# EXPLORING THE IMPACT OF SOCIAL STRESS ON THE ADAPTIVE DYNAMICS OF COVID-19: TYPING THE BEHAVIOR OF NAÏVE POPULATIONS FACED WITH EPIDEMICS

Innokentiy Kastalskiy[a,b,c,*], Andrei Zinovyev[d,e,f,#], Evgeny Mirkes[g], Victor Kazantsev[a,b,c], and Alexander N. Gorban[g]

[a]*Lobachevsky State University of Nizhny Novgorod, 23 Gagarin Ave., 603022 Nizhny Novgorod, Russia*
[b]*Moscow Institute of Physics and Technology, 9 Institutskiy Ln., 141701 Dolgoprudny, Moscow Region, Russia*
[c]*Samara State Medical University, 18 Gagarin St., 443079 Samara, Russia*
[d]*Institut Curie, PSL Research University, 75005 Paris, France*
[e]*INSERM U900, 75428 Paris, France*
[f]*Center for Computational Biology, Mines ParisTech, PSL Research University, 75006 Paris, France*
[g]*University of Leicester, University Rd, Leicester LE1 7RH, UK*
[#]*Current address: In silico R&D, Evotec, 31400 Toulouse, France*

*Corresponding author, e-mail: kastalskiy@neuro.nnov.ru

**ABSTRACT**
In the context of natural disasters, human responses inevitably intertwine with natural factors. The COVID-19 pandemic, as a significant stress factor, has brought to light profound variations among different countries in terms of their adaptive dynamics in addressing the spread of infection outbreaks across different regions. This emphasizes the crucial role of cultural characteristics in natural disaster analysis. The theoretical understanding of large-scale epidemics primarily relies on mean-field kinetic models. However, conventional SIR-like models failed to fully explain the observed phenomena at the onset of the COVID-19 outbreak. These phenomena included the unexpected cessation of exponential growth, the reaching of plateaus, and the occurrence of multi-wave dynamics. In situations where an outbreak of a highly virulent and unfamiliar infection arises, it becomes crucial to respond swiftly at a non-medical level to mitigate the negative socio-economic impact. Here we present a theoretical examination of the first wave of the epidemic based on a simple $SIR_{SS}$ model (SIR with Social Stress). We conduct an analysis of the socio-cultural features of naïve population behaviors across various countries worldwide. The unique characteristics of each country/territory are encapsulated in only a few constants within our model, derived from the fitted COVID-19 statistics. These constants also reflect the societal response dynamics to the external stress factor, underscoring the importance of studying the mutual behavior of humanity and natural factors during global social disasters. Based on these distinctive characteristics of specific regions, local authorities can optimize their strategies to effectively combat epidemics until vaccines are developed.

## 1. INTRODUCTION

The COVID-19 pandemic has emerged as a worldwide challenge for humanity, presenting a threat of this magnitude for the first time in modern history. As a result, there has been a significant increase in scientific research focused on coronaviruses and the spread of epidemics among populations. In addition to vaccination and other medical and biological interventions, theoretical research on the transmission of epidemics and the factors that influence their dynamics has become a crucial tool in the battle against infectious diseases.

Epidemics are one of the types of natural disasters that humanity can manage. In the dynamics of the natural disasters, the human response and the natural impact are intertwined, with societal reactions differing across countries. This distinction is where cultural characteristics play a pivotal role and must be taken into account in dealing with the disaster consequences.

The necessity to accurately explain the observed statistical patterns of viral epidemics and predict the impact of quarantine measures has driven researchers to resort to mathematical modeling. Various approaches have been employed, and among them, two contrasting ones stand out: the multi-agent approach [1-5], which describes the interactions among multiple agents in spatial contexts, and the mean-field approach [6-10], often summarized as "one like everyone else, and everyone like one". The compartmental models extend the predictive capabilities of the mean-field approach by adding compartments with different labels. People can move from one compartment to another. Inside the compartments, people are similar, but the difference between the compartments can be big.

The SIR model (Susceptible-Infected-Recovered) represents the classic and most logically simple compartmental population model with three compartments. Based on this framework, several modifications have been proposed to account for specific observed phenomena [11-15]. These kind of modifications and other related models include modeling the potential existence of a virus outside of carriers [16-18]; studying the role of information transmission within society [19,20]; the impact of control measures on epidemic dynamics [21-23]; and implementing spatially distributed models with pattern propagation on networks [24,25].

Periodicity is an important characteristic observed in many viral epidemics, indicating the presence of unstable attractors or oscillating solutions in the space of phase variables. The COVID-19 pandemic was no exception, as demonstrated by analyzing the statistics of SARS-CoV-2 infection cases and through theoretical approaches [26-30]. This periodicity can be attributed to several factors, including the seasonal decline of immunity, viral mutation leading to the emergence of new more virulent strains, re-infection among the affected population, lifting of restrictions by the authorities, or "exhaustion" of the population due to compliance with such measures, and other latent factors.

However, none of the arguments mentioned above can fully and comprehensively explain the real statistics of the epidemic throughout its entire duration. The initiation and spread of outbreaks within society's complex and multi-layered networks are influenced by many interconnected factors. These components work together in a way that cannot be neatly isolated or attributed solely to individual influences.

As part of the strategy to provide theoretical descriptions of COVID-19 dynamics, researchers often turn to modeling short epidemic outbreaks and/or focus on specific local regions, such as administrative districts [31,32]. It is well-known that dynamical systems are amenable to asymptotic analysis, and this applies to SIR-like models (for instance, see [33]).

In our previous work [34], we proposed a kinetic population model to study the spread of the COVID-19 epidemic outbreak (1). This model incorporates the influence of social stress, based on sociophysics methodologies [35], and allows multi-wave dynamics unlike the conventional SIR model. By combining a dynamic model resembling SIR with the classical triad of stages from the general adaptation syndrome (GAS) [36,37], we were able to accurately describe the available statistical data. Selye's discovery of GAS revealed the emergence of a typical syndrome with symptoms unaffected by the nature of the damaging agent. Integrated with Cannon's "fight-or-flight" response, this concept found diverse applications in physiology, psychology, and beyond [38-40], serving as essential models for our work.

GAS entails three chronological stages: *alarm*, *resistance*, and *exhaustion* [37]. After some rest the exhausted person can return to the initial state and become susceptible to alarm. Amid aggressive COVID-19 spread, the susceptible group can be segmented into three behavioral subgroups to replicate three types of population dynamics. We explore these behavioral modes based on the concept of GAS and social stress: 1) "ignorant mode" with no restrictions, 2) "resistance mode" following distancing and other protection rules, and 3) "exhaustion mode" leading to reduced adherence to follow social distancing rules. We consider a closed loop of transitions: ignorant → resistance → exhaustion → ignorant, where individuals return to the initial mode after exhaustion and become susceptible to alarm signals again. The resistant mode exhibits much lower infection probability than ignorant or exhaustion modes.

The conventional SIR model, enhanced by the SIR$_{SS}$ extension incorporating social stress, remarkably yielded excellent results in elucidating the influence of social stress on epidemic dynamics across diverse countries. Our model accommodated distinct outbreak profiles, acknowledging the substantial variations in social structures and cultural traditions. Interestingly, we found that these differences could be encapsulated in just a few kinetic constants that described the phases of social stress.

In the present work, we developed an analytical theory of the first wave of the epidemic by applying an asymptotic simplification to the SIR$_{SS}$ model. Within the new reduced model, we conducted an analysis to determine the influence of kinetic constants on the characteristics of the first surge. A cluster analysis was then performed on the resulting coefficients across all countries worldwide. This allowed us to identify distinctive patterns in the naïve population's behavioral response to the stress factor presented by the COVID-19 pandemic. Moreover, these patterns were found to roughly match the macro-regions to which the countries belonged.

## 2. MODEL AND METHODS USED

### 2.1. SIR$_{SS}$ model

The SIR-type model includes three variables reflecting fractions of the isolated population: susceptible (S), infected (I), and recovered (R). This simplest model depicts two possible reactions: $S + I \rightarrow 2I$ and $I \rightarrow R$. The reaction rate obeys the law of mass action, which sociophysics borrowed from chemical kinetics. However, these basic models fail to consider information and societal responses, e.g., hygiene and distancing decisions that impact epidemic transmission.

To address these gaps, following our previously published study [34], we introduced the refined model SIR$_{SS}$ taking into account the GAS theory. The stress response unfolds in three chronological stages: *alarm*, *resistance*, and *exhaustion*. Once an exhausted individual has had an opportunity to rest, they can return to their initial state and become susceptible to alarm once more. Assuming a fraction of the population is stress-free and neglecting the fleeting alarm stage, susceptible individuals (S) can be segmented into three subgroups based on their behaviors: those who are ignorant or unaware of the epidemic ($S_{ign}$), those who are rationally resistant ($S_{res}$), and those who are exhausted ($S_{exh}$) and are unresponsive to external stimuli (akin to a refractory period). In simple terms, this can be expressed as: $S(t) = S_{ign}(t) + S_{res}(t) + S_{exh}(t)$. This addition yields three new probabilistic transitions:

1) $S_{ign} + 2I \rightarrow S_{res} + 2I$ – rapid mobilization reaction reflecting an alarm response;
2) $S_{res} \rightarrow S_{exh}$ – slow exhaustion process due to COVID-19 restrictions;
3) $S_{exh} \rightarrow S_{ign}$ – gradual return to the initial state marking the end of the refractory period.

This refined model captures stress responses, resisting behavior exhaustion, and recovery to initial susceptibility. The autocatalytic form in the first reaction means that the transition rate is proportional to the square of the infected fraction I. The second corresponds to the depletion of the person's resources leading to reduction of following social distancing rules. The latter means that humans cannot be exhausted indefinitely.

The infection rates of $S_{exh}$ and $S_{ign}$ remain consistent with the basic SIR model, while the $S_{res}$ category transitions to I at a slower rate, considering this group as not susceptible to disease.

According to the described reactions, SIR$_{SS}$ model equations are as follows:

$$\frac{dS_{ign}}{dt} = -qS_{ign}I^2 - aS_{ign}I + K_3 S_{exh} \tag{1a}$$

$$\frac{dS_{res}}{dt} = qS_{ign}I^2 - K_2 S_{res} \tag{1b}$$

$$\frac{dS_{exh}}{dt} = -aS_{exh}I + K_2 S_{res} - K_3 S_{exh} \tag{1c}$$

$$\frac{dI}{dt} = aS_{exh}I + aS_{ign}I - bI \tag{1d}$$

$$\frac{dR}{dt} = bI \tag{1e}$$

where $a$ is morbidity rate, $b$ is recovery rate, $q$ is the stress response rate, $K_2$ is exhaustion rate, $K_3$ is the constant of transition to a state of ignorance are constant model parameters to be fitted from data. Possible transitions in the model are illustrated in **Fig. 1**.

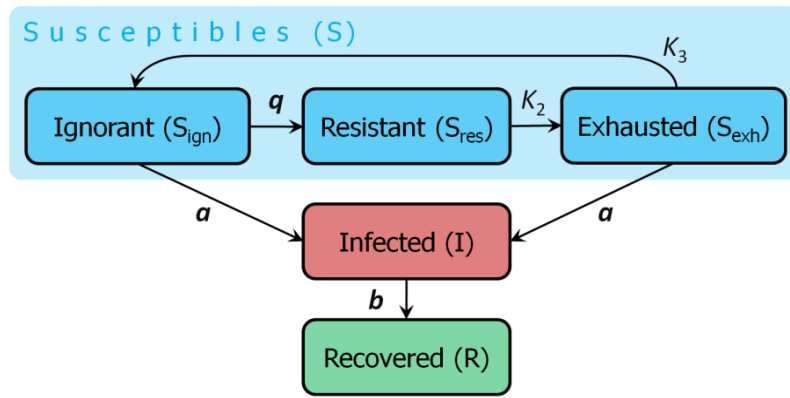

**Fig. 1.** A diagram of potential reactions within the $SIR_{SS}$ model with five compartments. Each arrow represents a specific kinetic constant. The "Susceptible" category exhibits a three-stage loop. Additionally, transition rates to the "Infected" class vary proportionally with the infected fraction I, while the rate of transition "Ignorant → Resistant" is proportional to $I^2$ (refer to the autocatalytic representation and (1) equations). Rapid transitions ($a$, $b$, $q$) are denoted in bold font, while slower transitions ($K_2$, $K_3$) are presented in regular font.

The $SIR_{SS}$ model was tested on datasets from 13 countries with high COVID-19 incidence rates. The studies demonstrated that by optimizing the model parameters, it was possible to accurately approximate the dynamics of the first and the onset of the second wave. Significantly, the resulting kinetic constants for individual countries/territories accurately captured the sociocultural characteristics of corresponding society. A detailed sociocultural analysis is beyond the scope of our work. However, we hope to provide a wealth of information for further analysis by the relevant expert community.

**2.2. Fitting the model to COVID-19 statistics**

In this study, we analyzed COVID-19 data for all countries worldwide. The computational modeling, model optimization, and comparison of obtained dynamics with observed data followed a similar approach to our previous paper [34]. We optimized model parameters for a specific population using the direct fitting method: we calculated the model's dynamics and compared each resulting curve with the observed COVID-19 statistics. The correspondence was estimated using the coefficient of determination $R^2$. The search for the global maximum of the $R^2$ was implemented using a uniform grid in the parameter space ($a$, $K_2$, $q$, $I_0$), at which the model reproduces epidemics dynamics that are closest to the observed one.

The primary kinetic constants were $a$ (morbidity rate), $q$ (stress response rate), and $K_2$ (exhaustion rate). These constants were supplemented with statistics on GDP per capita at PPP (gross domestic product per capita at purchasing power parity) for 2020, as well as minor values characterizing the specific implementation of the model: $I_0$ (initial fraction of infected population) and $N_{days}$ (number of modeled days). The parameters $b$ (recovery rate) and $K_3$ (slow constant of transition to a state of ignorance) were fixed, as explained in the previous work [34].

**2.3. Statistical processing and cluster analysis**

As mentioned earlier, the model parameters determine the epidemic profiles for each country. Furthermore, it was observed that neighbouring countries with similar economical, historical, ethnic, linguistic, social, and cultural features exhibited highly similar COVID-19 dynamics. This observation allowed us to group these countries into six large historically developed territorial and cultural-economic regions: (i) Africa (excluding North Africa); (ii) North Africa and the Middle East; (iii) Asia (excluding the Middle East and the Asian part of the Russian Federation); (iv) Central and South America; (v) Western Europe, USA and three countries of the British Commonwealth – Canada, Australia, New Zealand; (vi) Eastern Europe (including the Asian part of the Russian Federation). We recognize the likelihood of discussions and disagreements concerning the regional divisions, which were partially formed based on the authors' subjective assessment of current COVID-19 data throughout the study.

To verify and to visualize the hypothesis of the existence of distinct social groups exhibiting different dynamics in response to stress factors like the COVID-19 pandemic, we applied the elastic maps method of non-linear dimensionality reduction [41-46]. This method computes an explicit representation of two-dimensional principal manifold that approximates the finite multivariate data point cloud in the mean-square error sense and simultaneously possesses regular properties such as smoothness. Thanks to this explicit manifold representation the multidimensional data vectors can be projected onto the manifold together with smoothed nonlinear trends present in the data for further data visualization.

In addition, we applied several clustering methods, including k-means (with 1000 runs for random initial centroids), agglomerative, spectral, DBSCAN, and hierarchical clustering using Ward linkage. The adjusted Rand index (ARI) and silhouette score were used as measures to assess the quality of clustering.

## 3. THEORY

Let us consider an analytical study of the dynamics during the first wave of the coronavirus in the $SIR_{SS}$ model. Research has demonstrated that outbreaks occur on relatively small time scales, during which the influence of the terms $K_2 S_{res}$, $K_3 S_{exh}$, and $a S_{exh} I$ on the right-hand side of the equations (1a-1d) can still be considered negligible. This led us to focus our study on the $SIR_{SS}$ model specifically during the initial surge of the coronavirus epidemic.

### 3.1. Phase portrait I(S) of the first wave of the $SIR_{SS}$ COVID-19 model

We can rewrite equations (1a) and (1d) in the following form:

$$\frac{dS_{ign}}{dt} = -q S_{ign} I^2 - a S_{ign} I \tag{2a}$$

$$\frac{dI}{dt} = a S_{ign} I - b I \tag{2b}$$

Next, we will construct an analytical phase portrait in the ($S_{ign}$, I) axes. To achieve this, we divide equation (2a) by equation (2b):

$$\frac{dS_{ign}}{dI} = \frac{S_{ign}(a + qI)}{b - a S_{ign}} \tag{3}$$

Having solved the differential equation above using the separation of variables method, we substitute the initial condition $I_0 + S_0 = 1$ and make the substitutions $S \to S_{ign}$, $b/a \to b_a$, and $a/q \to a_q$ for readability. As a result, we will have:

$$I_{SIR_{SS}}(S) = -a_q + \sqrt{a_q^2 + I_0^2 + 2a_q\left(1 - S + b_a \ln\frac{S}{S_0}\right)} \tag{4}$$

For comparison, the phase portrait of a standard SIR:

$$I_{SIR}(S) = 1 - S + b_a \ln \frac{S}{S_0} \quad (5)$$

Note that it is necessary to take into account the existence of the remaining three variables: $S_{res} \geq 0$, $R \geq 0$, and $0 \leq S_{exh} \ll 1$.

**Fig. 2** represents the most general form of the phase portrait for the reduced SIR$_{SS}$ model. Looking ahead, it is worth mentioning that the shapes of the trajectories are qualitatively the same for both models and under variations in key parameters. The differences lie primarily in quantitative indicators, which will be discussed in the survey below. The epidemic will spread over a certain period of time, contingent upon the initial condition $I_0$, and will eventually reach its peak. As the time increases, our assumptions will gradually become less effective, causing the reduced model to diverge further from the original SIR$_{SS}$ model.

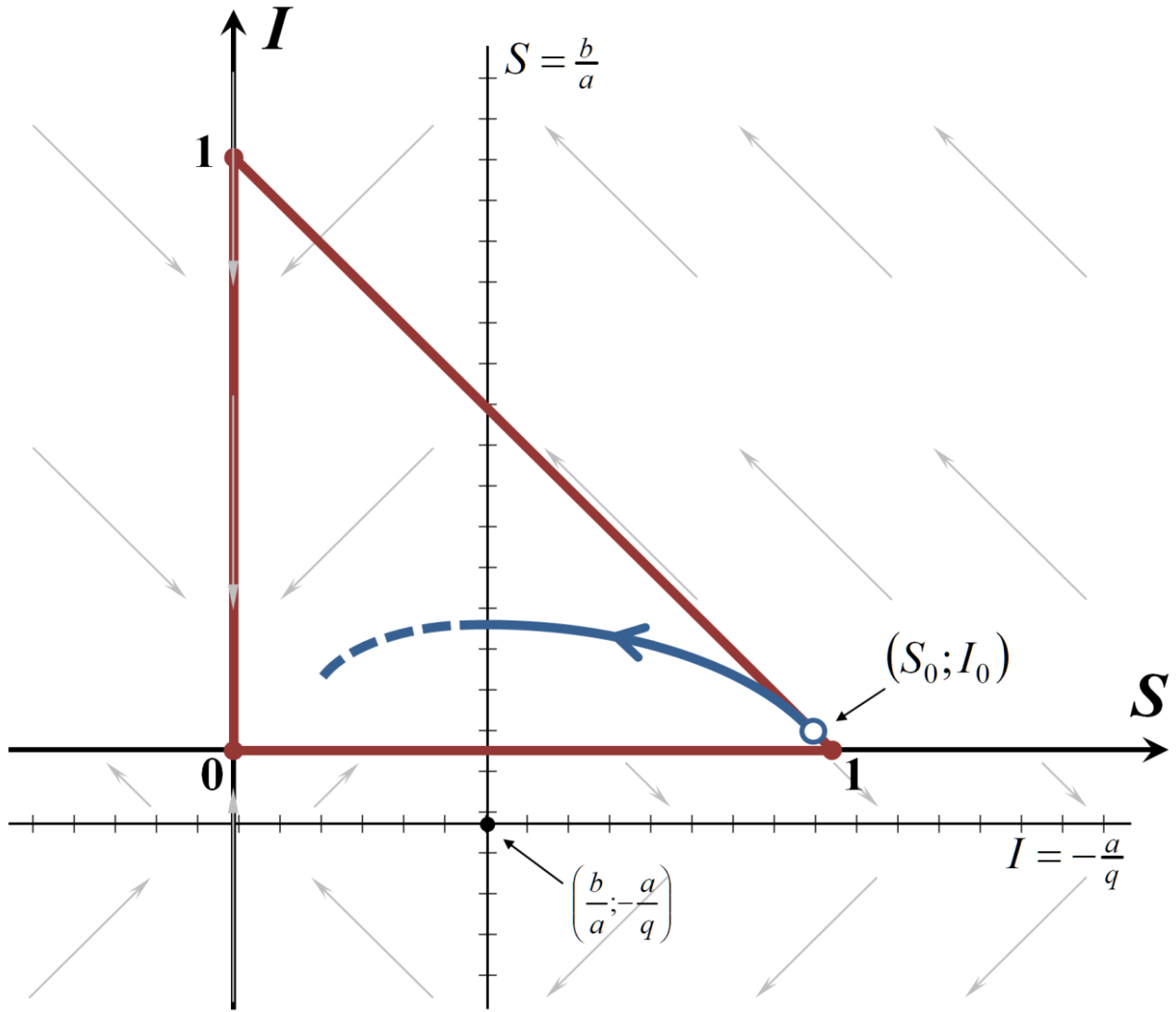

**Fig. 2.** Phase space for the SIR$_{SS}$ model (*for clarity, the scales are not fully respected!*). There is only one steady state: $(b_a; -a_q)$, which is located outside the working area of the model. The real dynamics of the system is possible only inside the triangle with vertices (0;0), (1;0), (0;1) – colored in red online, because the following conditions are always satisfied: $S, I \geq 0$, $S + I \leq 1$. Lines with multiple dashes correspond to two major isoclines, while gray arrows depict the vector field indicating the directions of a moving flow of the variables in phase space. The degenerate steady line $I = 0$ changes its stability when it intersects with the nullcline $S = b_a$.

Starting from the initial conditions $(S_0; I_0)$ on the line $I = 1 - S$, located near the point (1;0), we will cross the isocline $S = b_a$ horizontally (this will correspond to the peak of the first wave) and will begin to be attracted to the stable part of $I = 0$.

The amplitude of the first wave can be estimated analytically:

$$I_{max,SIR_{SS}} = I(b_a) = -a_q + \sqrt{a_q^2 + I_0^2 + 2a_q\left(1 - b_a + b_a \ln \frac{b_a}{1 - I_0}\right)} \qquad (6)$$

In comparison, for a conventional SIR model:

$$I_{max,SIR} = 1 - b_a + b_a \ln \frac{b_a}{1 - I_0} \qquad (7)$$

These two dependencies are shown in **Fig. 3A**. Here and below in numerical calculations we will use the parameters determined for Russian Federation (unless otherwise specified): $a = 0.1495$, $b = 0.1$, $q = 47500$, $I_0 = 3.59 \times 10^{-5}$. For the limiting case of an extremely small $I_0$, we get:

$$I_{max,SIR_{SS}} = -a_q + \sqrt{a_q^2 + 2a_q(1 - b_a + b_a \ln b_a)} \approx 0.0006222 \qquad (8)$$

$$I_{max,SIR} = 1 - b_a + b_a \ln b_a \approx 0.06212$$

Thus, the amplitudes of the peaks $I_{max}$ for the models differ by almost two orders of magnitude (**Fig. 3B**).

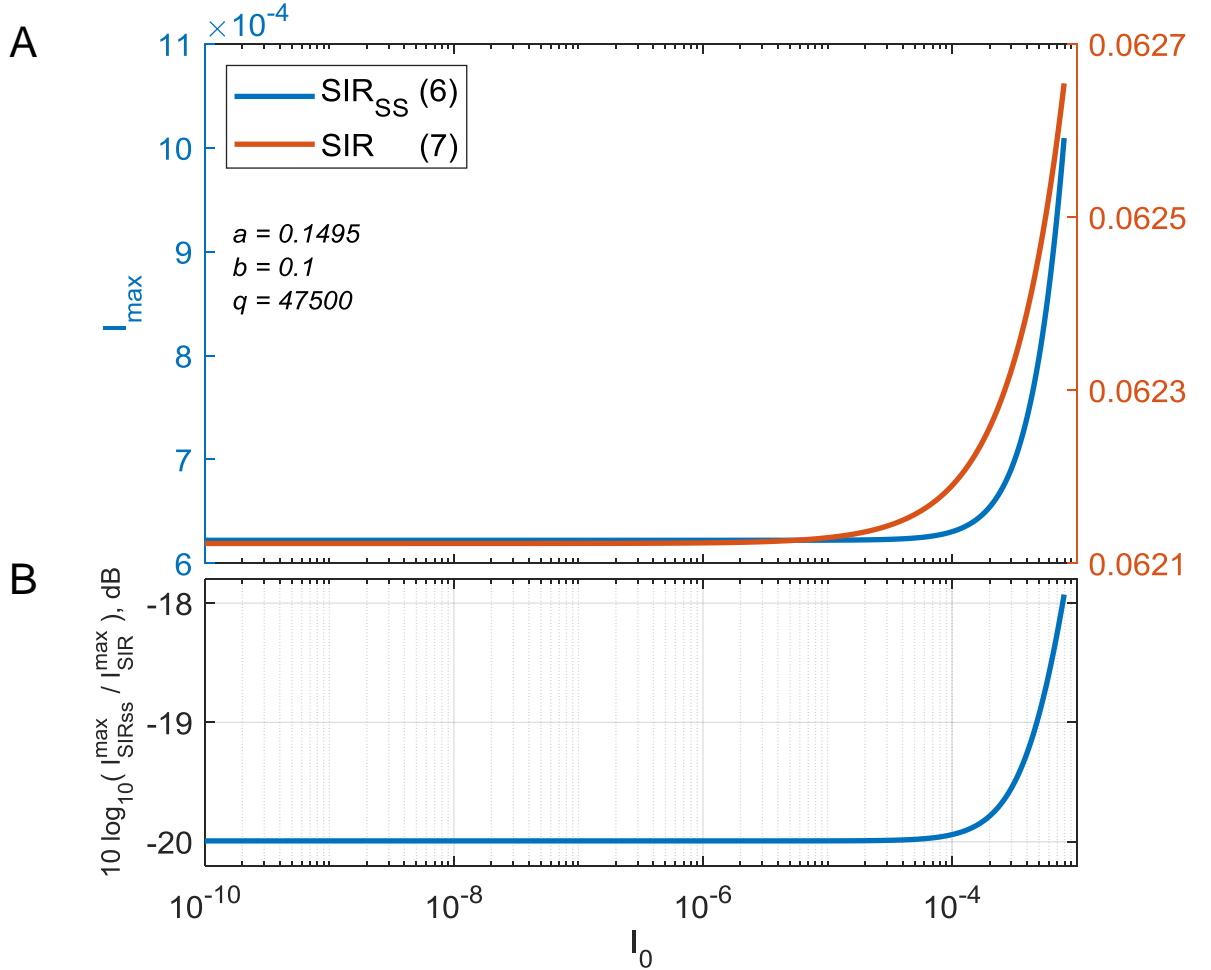

**Fig. 3. A** – The amplitude of the first wave $I_{max}$ for the models $SIR_{SS}$ (6) and SIR (7) depending on the initial fraction of infected population $I_0$. It is obvious that for $I_0 < 10^{-4}$ the amplitude of the first wave peak is constant: $I_{max} = 6.222 \times 10^{-4}$. The only difference is the time it takes to reach the peak. In the case of a conventional SIR for the investigated range $I_0 \in [10^{-10}; 10^{-3}]$ $I_{max}$ changes insignificantly and is set at the level of 0.06212 for $I_0 < 10^{-5}$; **B** – The ratio of the amplitudes of the reduced $SIR_{SS}$ and the standard SIR model. For $I_0 < 10^{-4}$, the value is -20 dB.

**Fig. 4** shows phase portraits for the SIR$_{SS}$ model for $I_0 = 10^{-10}$, $1\times10^{-4}$, $2\times10^{-4}$, and $5\times10^{-4}$:

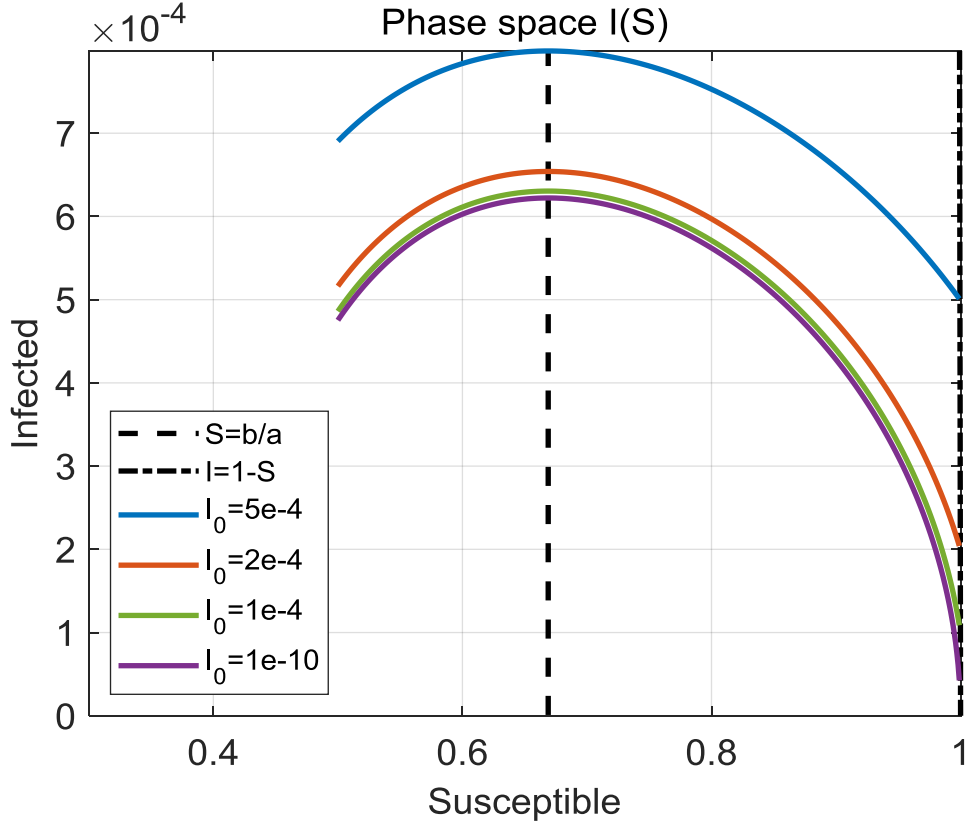

**Fig. 4.** Phase portraits I(S) for different values $I_0 = 10^{-10}$, $1\times10^{-4}$, $2\times10^{-4}$, and $5\times10^{-4}$. The dashed line indicates the nullcline, the dash-dotted line indicates the line $I = 1 - S$. It can be seen that all curves I(S) that meet the condition: $I_0 < 10^{-4}$ will repeat each other, i.e. for actually observed initial conditions, when $I_0$ is very small, the dynamics will be almost the same.

When compared with the conventional SIR model, it was found that the curves are quite similar to SIR$_{SS}$, but the trajectories are at least two orders of magnitude higher for $I_{SIR_{SS}} > 6.25\times10^{-4}$:

$$I_{SIR_{SS}}(S) = -a_q + \sqrt{a_q^2 + I_0^2 + 2a_q I_{SIR}(S)} \approx \sqrt{2a_q I_{SIR}(S)} \Rightarrow$$

$$\Rightarrow \frac{\sqrt{I_{SIR}}}{I_{SIR_{SS}}} \approx \sqrt{\frac{q}{2a}} \approx 400 \quad and \quad I_{SIR} = 160000\, I_{SIR_{SS}}^2 \Rightarrow \qquad (9)$$

$$\Rightarrow \quad if\ I_{SIR_{SS}} > 1/1600 = 0.000625,\ then\ \frac{I_{SIR}}{I_{SIR_{SS}}} > 100.$$

We would like to highlight here the following observations:
1) the shape of the phase trajectories of the enhanced and then simplified dynamic model, SIR$_{SS}$, is not sensitive to the parameter $I_0$;
2) phase portraits of the SIR$_{SS}$ model exhibit nearly identical patterns for any $I_0 < 10^{-4}$;
3) the inclusion of feedback on $I^2$ in the SIR model significantly reduces the amplitude of the first wave peak by two orders of magnitude; for example, this is the case of the computed kinetic constants previously determined for Russian Federation.

### 3.2. Influence of parameters *a* (morbidity rate) and *q* (mobilization rate)

Let us now study the effect of the parameter *a* on the peak amplitude. **Fig. 5A** shows two dependencies $I_{max}(a)$ for the SIR$_{SS}$ and SIR models. Amplitude ratio $I_{max,SIRss}/I_{max, SIR}$ is set at -22 dB for $a > 0.2$ (**Fig. 5B**).

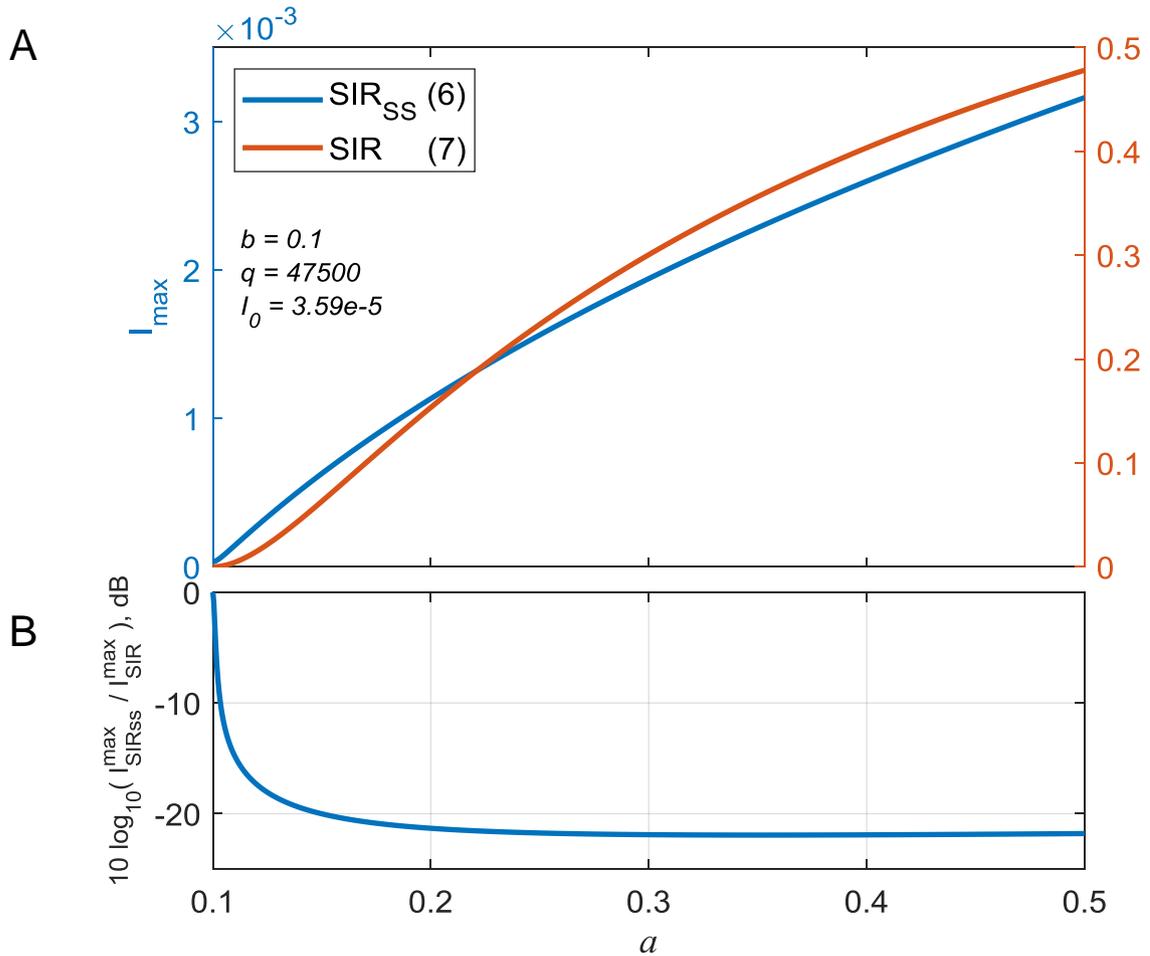

**Fig. 5. A** – Dependences of the first epidemic wave peaks on the morbidity rate *a*. For $a < b = 0.1$, the epidemic outbreak does not progress at all and $I_{max} = I_0$. For $a > b$, the dependences for both models are quasi-linear and closely repeat each other in shape. The difference in the scales of the vertical axes indicates a difference in values by two orders of magnitude; **B** – The ratio of the amplitudes of the enhanced SIR$_{SS}$ and the standard SIR model. For $a > 0.2$, the value is -22 dB.

The phase portraits for SIR$_{SS}$ will differ only in the shape of the nullcline $S = b_a$, henceforth it is not a line, but looks like a hyperbole (**Fig. 6**):

$$I(S) = -a_q + \sqrt{a_q^2 + I_0^2 + 2a_q\left(1 - S + b_a \ln\frac{S}{S_0}\right)} \xrightarrow{a=\frac{b}{S}} \approx \sqrt{\frac{2b}{q}\left(\frac{1}{S} + \ln\frac{S}{1 - I_0} - 1\right)} \quad (10)$$

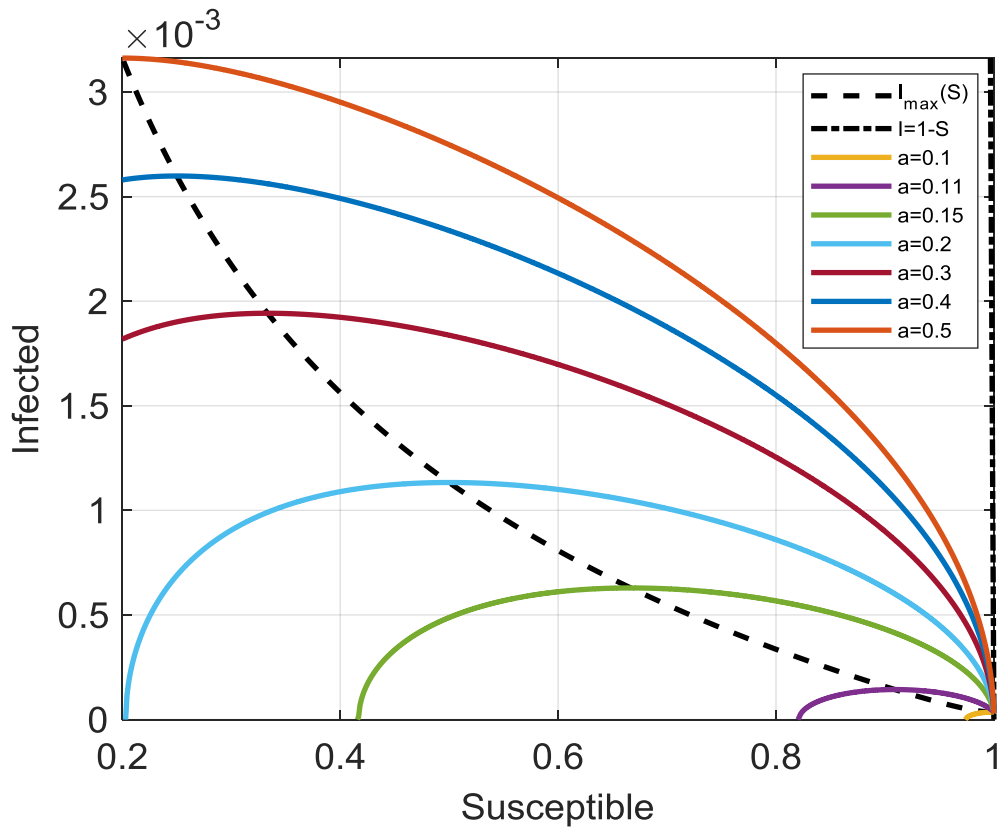

**Fig. 6.** Phase portraits I(S) for different values $a$ = 0.1, 0.11, 0.15, 0.2, 0.3, 0.4, and 0.5. The dashed line indicates the nullcline, the dash-dotted line indicates the line I = 1 – S.

Let us consider the influence of the parameter $q$. For countries from South America $q \in [10^3; 10^4]$, for ordinary European countries $q \in [10^4; 10^5]$, for countries capable of rapid mobilization $q \in [10^5; 10^6]$, for extremely regulated ones, such as Iran and China, $q > 10^6$. **Fig. 7A** illustrate how the $q$ parameter affects the maximum amplitude of the first wave $I_{max}$. The ratio of the amplitudes of the first wave illustrated on the **Fig. 7B**.

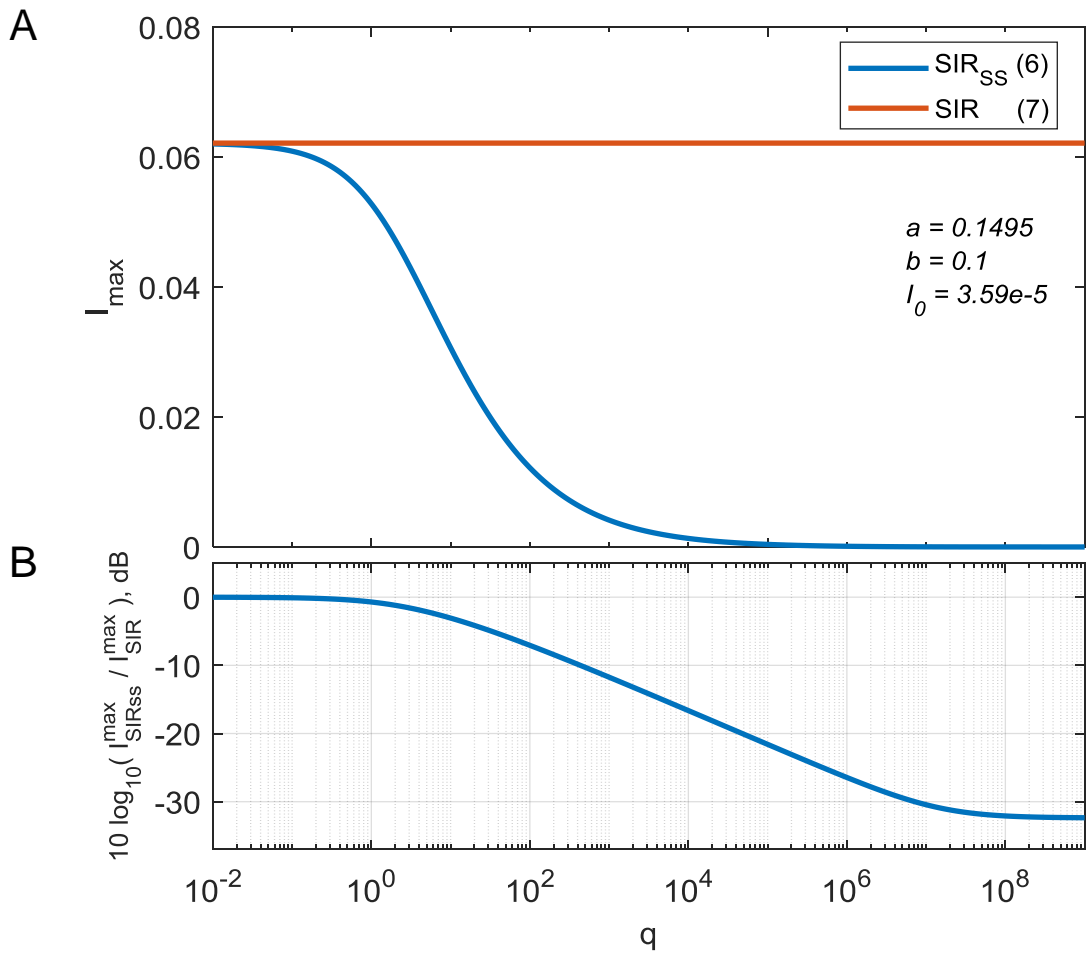

**Fig. 7. A** – Dependences of the first epidemic wave peaks on the mobilization rate $q$. The dependence is a sigmoid-like function with a midpoint $q \approx 10$. It can be seen the parameter reflects the difference between the $SIR_{SS}$ model and the classical SIR, for $q = 0$ simplified $SIR_{SS}$ become the same as SIR; **B** – The ratio of the amplitudes of the enhanced $SIR_{SS}$ and the standard SIR model. For actually observed values of the parameter $q > 10^3$, the amplitude of the first peak of the epidemic dramatically falls by more than one order of magnitude. For $q > 10^8$, the value is -32 dB.

A series of phase portraits I(S) for SIR$_{SS}$ model is displayed in **Fig. 8**.

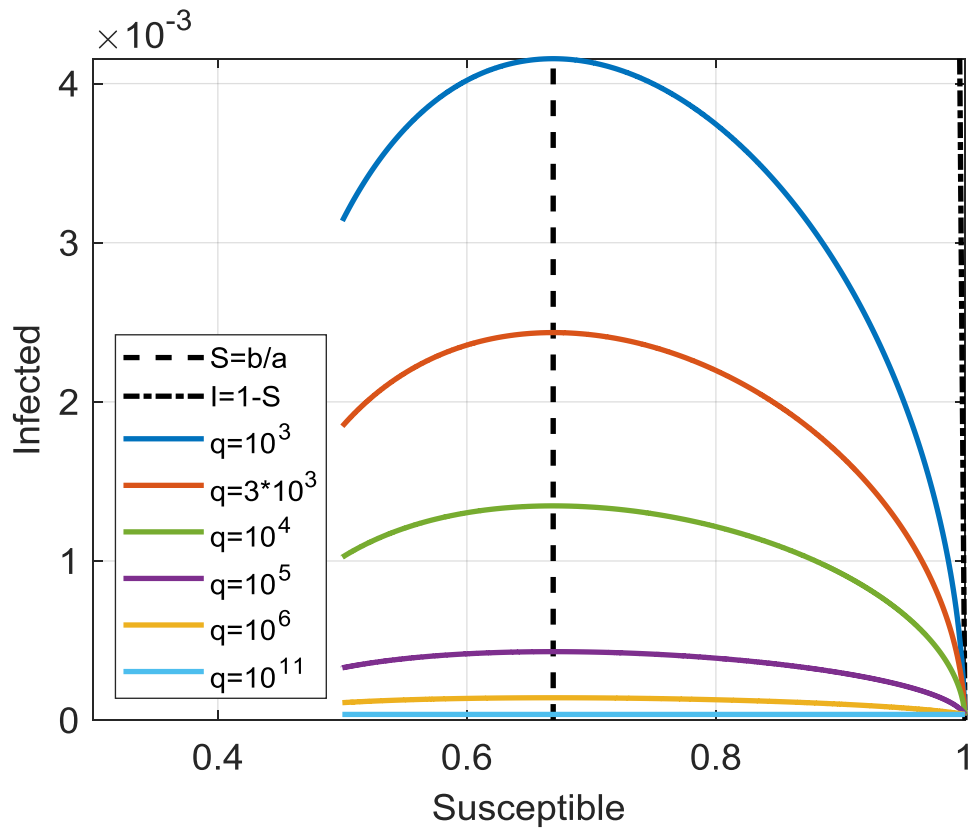

**Fig. 8.** Phase portraits I(S) for different values $q = 10^3$, $3 \times 10^3$, $10^4$, $10^5$, $10^6$, and $10^{11}$. The curves have similar shape and differ only in the amplitude $I_{max}$ at the peak of the outbreak. One of these curves at $q = 0$ will correspond to the trivial SIR model (not shown), the amplitudes of the rest decrease with increasing $q$.

### 3.3. The problem of parameter *b* (recovery rate)

The duration of infection for a susceptible individual is much shorter than the overall duration of the epidemic. In the SIR model, the coefficient *a* represents the efficiency of infecting others (in terms of the reciprocal of the characteristic time), while the constant *b* reflects the reciprocal of the effective time interval from infection to isolation of the individual. Consequently, using short periods such as two days (with a value of $b = 0.5$) would imply an impractically rapid transition of people to self-isolation or hospitalization on a state-wide scale. In light of these considerations, we decided to fix the value of $b = 0.1$ (equivalent to 10 days) to allow for an optimization of parameter *a* within manageable ranges. As we have verified earlier, it is the ratio of $b/a$ that primarily influences the dynamics of the first wave of the epidemic, as demonstrated by formulas (4) and (6). Therefore, we can anticipate that the influence of *b* (with *a* and *q* being fixed) will be directly opposite to that of *a*. Let us study the dependence of the phase portrait on the recovery rate *b*. Wave amplitude $I_{max}$ is shown on the **Fig. 9A**. The amplitude ratio $I_{max,SIRss}/I_{max,SIR}$ is displayed on the **Fig. 9B**.

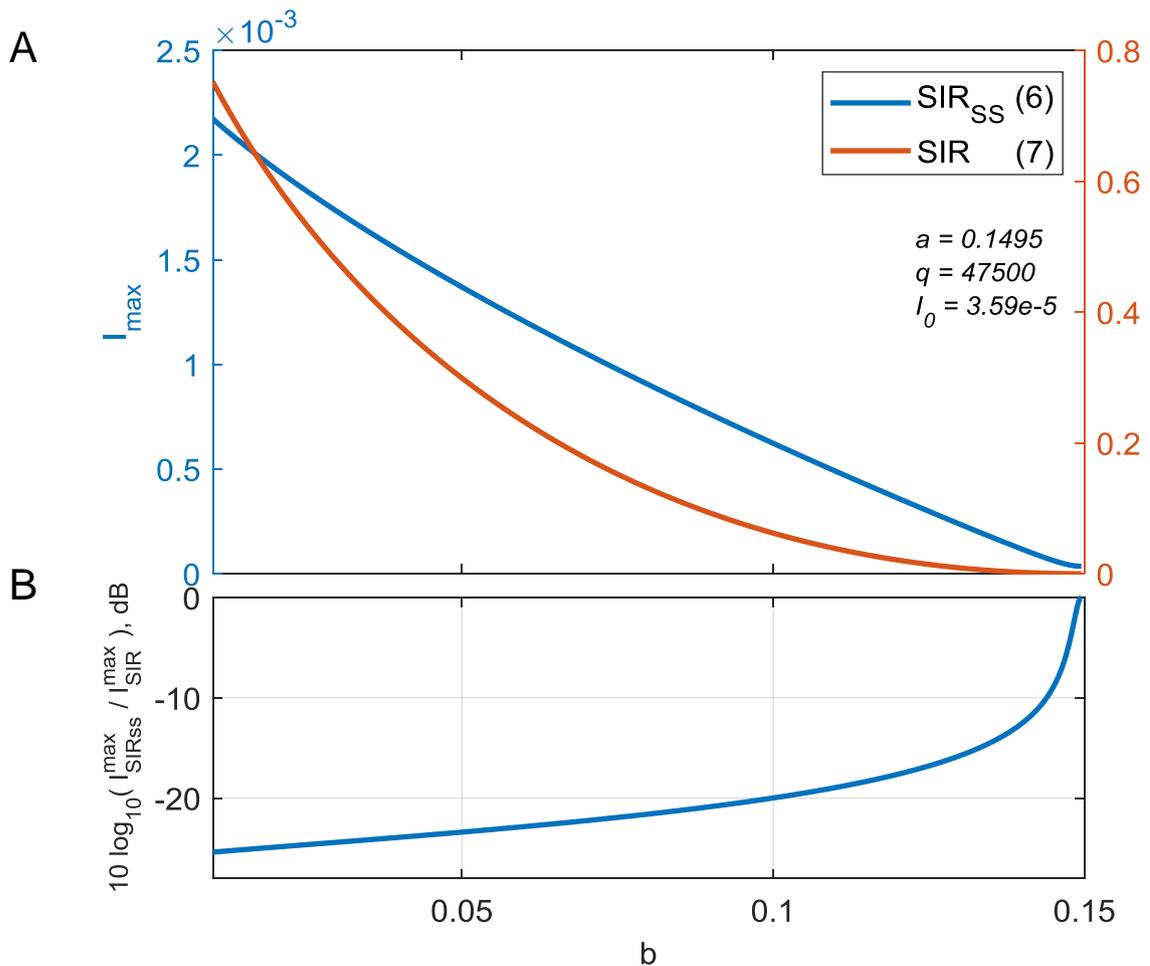

**Fig. 9. A** – Dependences of the first epidemic wave peaks on the recovery rate *b*. For $b > a = 0.1495$, the epidemic outbreak does not progress at all and $I_{max} = I_0$. For $b < a$, the dependency for the SIR$_{SS}$ model is quasi-linear. The difference in the scales of the vertical axes indicates a difference in values by two orders of magnitude; **B** – the ratio of the amplitudes of the enhanced SIR$_{SS}$ and the standard SIR model

Phase portraits for a series of 7 different *b* in the interval [0.025; 0.15] is plotted on **Fig. 10**.

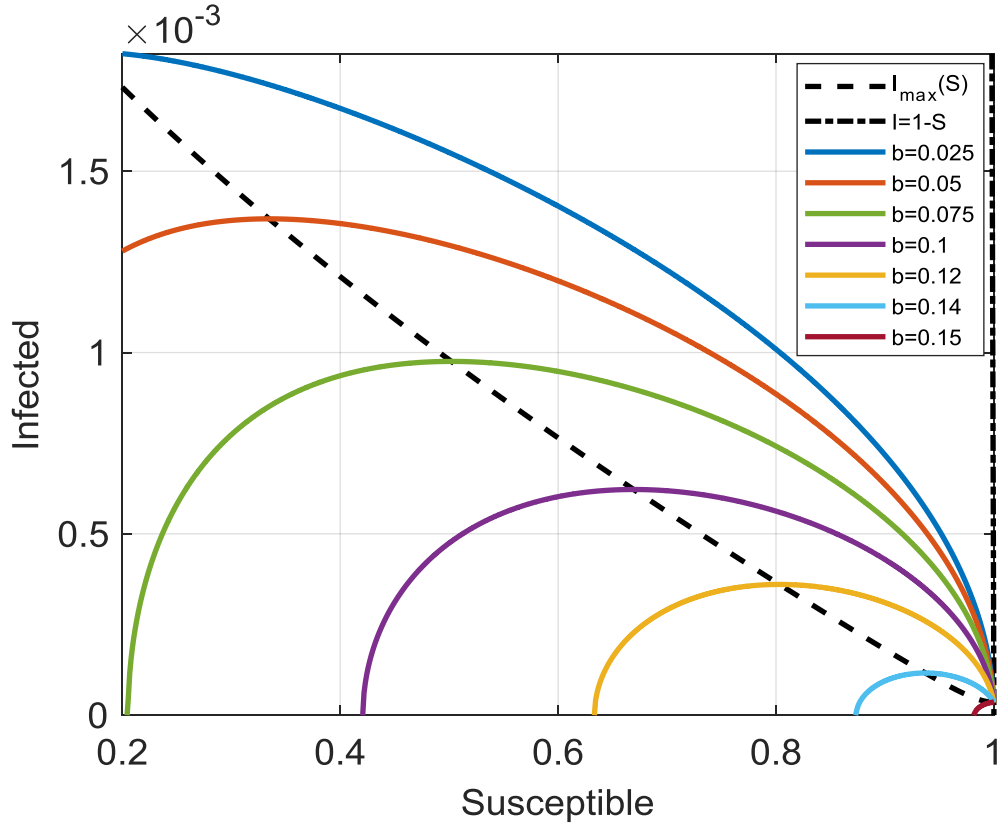

**Fig. 10.** Phase portraits I(S) for SIR$_{SS}$ model for different values *b* = 0.025, 0.05, 0.075, 0.1, 0.12, 0.14, and 0.15. The nullcline is almost linear.

Similar to the variation in the morbidity rate *a* relative to the recovery rate *b*, a change in parameter *b* relative to *a* results in a shift in the peak position of the epidemic wave in the phase space.

We can make the following conclusions:
- adding the term $qSI^2$ to S equation in the SIR model leads to a decrease in the amplitude of the epidemic wave by orders of magnitude compared to the original SIR model with the same parameters;
- modifying the parameters *a*, *b*, *q*, and the initial condition $I_0$ does not cause qualitative changes in the phase portrait;
- the mutual ratio of *b* and *a* determines the incidence rate, the amplitude of the first peak and its position in the phase space (S, I).

## 4. RESULTS AND DISCUSSION

### 4.1. Properties of the first wave of the COVID-19 epidemic

Based on the analytical theory of the first wave using a specific set of kinetic constants, examining the system's dynamics in the phase plane I(S$_{ign}$) (**Fig. 2**), the following properties of the SIR$_{SS}$ model were identified:
- the incorporation of feedback on $I^2$, influenced by the social stress factor, significantly reduces the peak of the first wave compared to the original SIR model with identical parameters;
- variations in the parameters *a*, *b*, *q*, and the initial condition $I_0$ do not result in qualitative changes in the phase portrait. Moreover, the phase trajectories are almost identical for any $I_0 < 10^{-4}$ and other parameters being fixed;
- the mutual ratio of *b* and *a* determines the COVID-19 incidence rate, the amplitude of the first peak, and its position in the phase plane I(S$_{ign}$).

In other words, the analytical theory (see Section 3) predicts similar outbreak scenarios during the initial stage of an epidemic, where society has not yet adapted to the global changes in the epidemiological landscape. The variability in observed COVID-19 statistics across countries is primarily determined by quantitative variations in the kinetic constants that are characteristic of specific countries, territories, social groups, etc.

### 4.2. Analysis of all countries worldwide

To conduct a comprehensive analysis, we first filtered out countries and territories from the available 209 for analysis, resulting in 169 subjects with sufficient statistical data. A fragment of the results of fitting parameters of model (1) is presented in Table 1. For reference, the profiles of COVID-19 outbreak dynamics in each country are provided in Appendix A (**Fig. A.1** and **Fig. A.2**).

Table 1. Kinetic constants fitted for each country of the world from observed epidemic dynamics (*fragment*)

| ISO 3 | ISO 2 | Continent | Region | Population | Threshold | $N_{days}$ | Country | GDP per capita PPP | $a$ | $K_2 \times 10^{-3}$ | $q \times 10^3$ | $I_0 \times 10^{-6}$ | $R^2$ |
|---|---|---|---|---|---|---|---|---|---|---|---|---|---|
| AFG | AF | Asia | Asia excl. Middle East | 38 928 341 | 100 | 250 | Afghanistan | 2078,6 | 0,1791 | 5,30 | 1440 | 1,65 | 0,99868 |
| ALB | AL | Europe | Eastern Europe | 2 877 800 | 10 | 100 | Albania | 13439,7 | 0,2117 | 20,61 | 22295 | 5,36 | 0,99311 |
| DZA | DZ | Africa | North Africa & Middle East | 43 851 043 | 10 | 275 | Algeria | 11324,2 | 0,1311 | 20,22 | 1328 | 2,85 | 0,98869 |
| AGO | AO | Africa | Africa excl. North | 32 866 268 | 100 | 325 | Angola | 6445,4 | 0,1333 | 7,63 | 4025 | 1,18 | 0,99735 |
| ARG | AR | South America | Central & South America | 45 195 777 | 100 | 300 | Argentina | 20770,7 | 0,1283 | 9,61 | 0,96 | 22,6 | 0,99953 |
| ARM | AM | Asia | Eastern Europe | 2 963 234 | 100 | 200 | Armenia | 13312,1 | 0,1509 | 8,97 | 6,09 | 30 | 0,99934 |
| AUS | AU | Oceania | Western Europe & USA & British Cmlth | 25 499 881 | 10 | 175 | Australia | 53329,8 | 0,2858 | 7,24 | 24161 | 0,01 | 0,97934 |
| AUT | AT | Europe | Western Europe & USA & British Cmlth | 9 006 400 | 100 | 200 | Austria | 55683,8 | 0,2942 | 5,25 | 460,6 | 21 | 0,99496 |
| AZE | AZ | Asia | Eastern Europe | 10 139 175 | 10 | 225 | Azerbaijan | 14479,7 | 0,1503 | 7,46 | 91,6 | 3,25 | 0,99866 |
| BHS | BS | North America | Central & South America | 393 248 | 100 | 350 | Bahamas | 32538,6 | 0,1441 | 5,99 | 3,36 | 21,7 | 0,99600 |
| BHR | BH | Asia | North Africa & Middle East | 1 701 583 | 10 | 200 | Bahrain | 43755,9 | 0,1443 | 21,23 | 2,21 | 35,6 | 0,99973 |
| BGD | BD | Asia | Asia excl. Middle East | 164 689 383 | 100 | 250 | Bangladesh | 5138,7 | 0,1394 | 10,89 | 330,6 | 11,5 | 0,99943 |
| BLR | BY | Europe | Eastern Europe | 9 449 321 | 10 | 250 | Belarus | 20239,2 | 0,1838 | 7,45 | 31,3 | 10,2 | 0,99566 |
| BEL | BE | Europe | Western Europe & USA & British Cmlth | 11 589 616 | 100 | 200 | Belgium | 53035 | 0,2153 | 7,10 | 58 | 36 | 0,99701 |
| … | … | … | … | … | … | … | … | … | … | … | … | … | … |

Secondly, we have deemed ($a$, $q$, $K_2$, GDP) as robust measures of country characteristics because only the first three reliably determine the outbreak profile, while GDP is widely recognized as an indicator of economic prosperity for countries.

Thirdly, we estimated the data scatter for each parameter. It was observed that the kinetic constant $q$ has an extremely uneven distribution, suggesting the use of a logarithmic scale, $\log(q)$, for further studies. Additionally, four countries: Vietnam, Timor, Taiwan, and Uganda, exhibited exceptionally high values (outliers) with $q > 10^9$. As a result, it was decided to exclude these countries when constructing the elastic maps and performing further data classification.

The remaining 165 countries formed a multidimentional data point cloud whose intrinsic structure was studied by computing 2D principal manifold, using the elastic map method [41-46], see **Fig. 11**. We projected the data from the 4D space onto the principal manifold and estimated the density of the projections in 2D (**Fig. 12A**). Furthermore, the continuous smoothed values for each of the mentioned features, namely ($a$, $\log(q)$, $K_2$, GDP), were plotted to represent their smoothed trends on the data map (**Fig. 12B-E**, respectively). The color scales in these figures represent the feature values, ranging from the smallest (pale colors) to the largest (saturated colors).

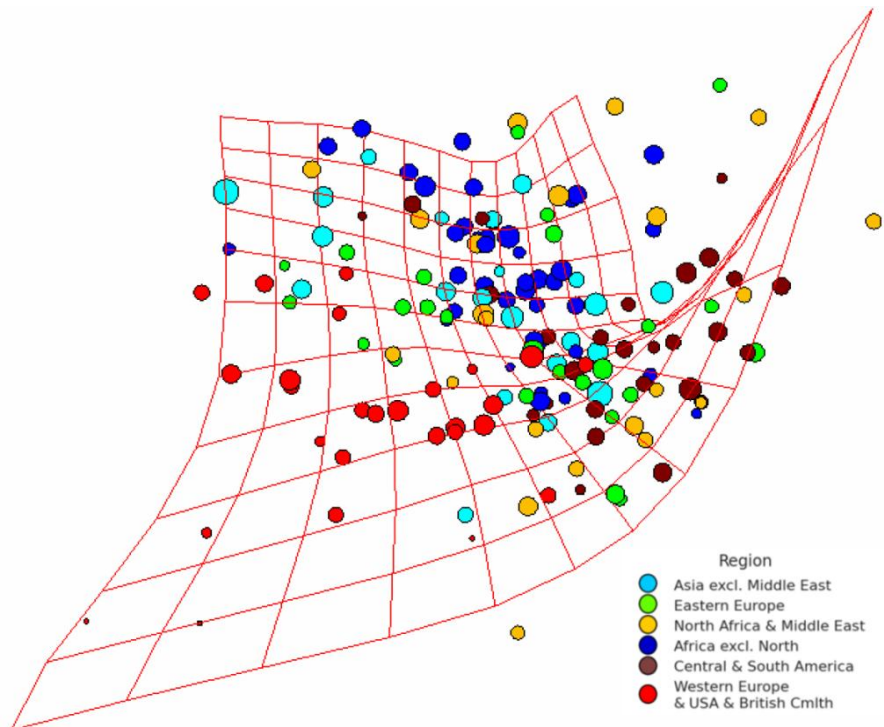

**Fig. 11.** Elastic map superimposed on projections of data in the space of the first three principal components. Colors encode regions grouped according to economic and socio-cultural features into 6 regions (see Section 2.3 for more details). The sizes of the points are logarithmically proportional to the population size of the corresponding country.

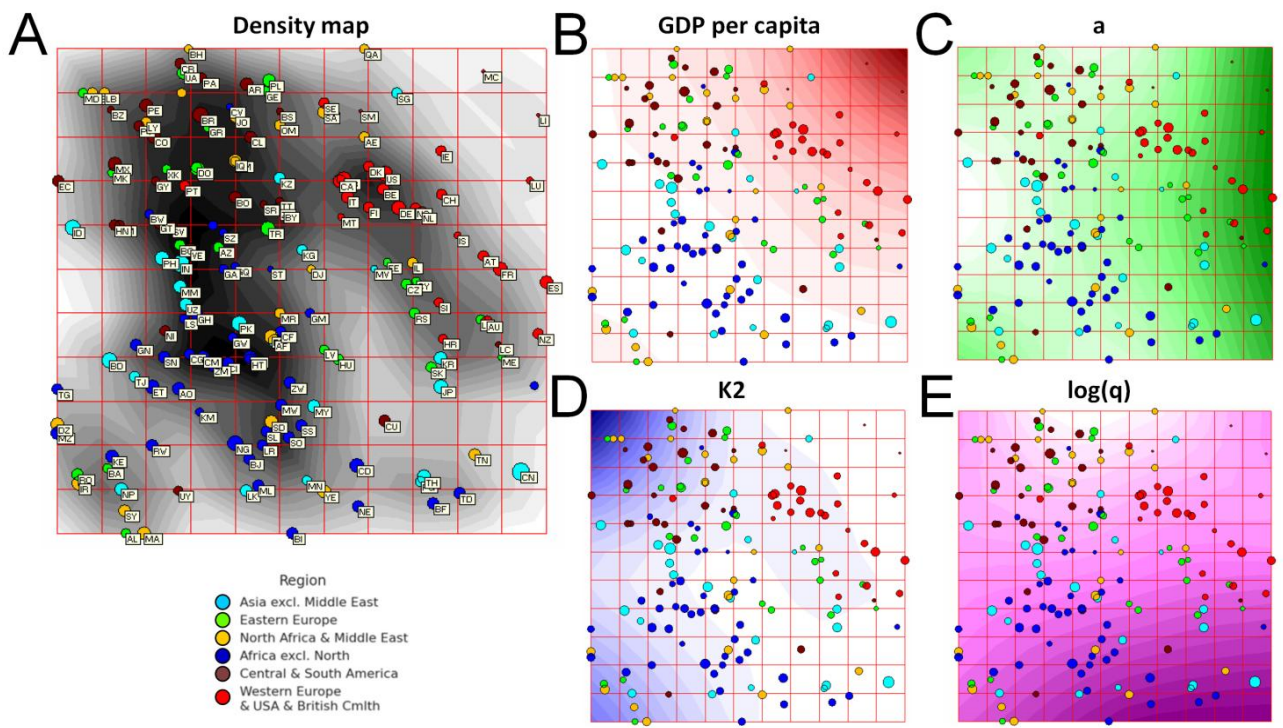

**Fig. 12.** The 2D map represents the distribution of projections onto the manifold shown in **Fig. 11**, identical in all panels except the background coloring. Coordinate and point density colorings of the map of kinetic constants are shown, including: the total density of point projections (**A**), GDP per capita at PPP (**B**), morbidity rate $a$ (**C**), exhaustion rate $K_2$ (**D**), logarithm of stress response rate $q$ (**E**). More intensive color of the background corresponds to greater values. The individual point coloring and sizes are the same as in **Fig. 11**. Countries are labeled using their standard ISO Alpha 2 two-letter abbreviations.

Observing the density distributions of countries on the elastic map (**Fig. 12**) reveals several interesting properties:
1. Many countries form one or two dominant clusters in the left part of the elastic map, from top to bottom (**Fig. 12A**). This point agglomeration corresponds to regions with generally low morbidity rates (**Fig. 12C**, green). In addition, few tens of countries form a cluster in the middle of the right part of the map, following a diagonal shape. This cluster is distinctive by high GDP (**Fig. 12B**, red). Furthermore, a cluster is observed in the lower left corner, where high exhaustion (**Fig. 12D**, blue) and mobilization constants (**Fig. 12E**, magenta) are observed.

2. Central and South American countries group in the upper left corner, where they are connected to two other regions: Eastern Europe and the Middle East and North Africa. As a result, this region of the elastic map is characterized by the highest density (**Fig. 12A**). Notably, this quadrant also displays relatively low log($q$) values (**Fig. 12E**, magenta), and high values of the K$_2$ parameter (**Fig. 12D**, blue).
3. Sub-Saharan Africa and a significant portion of Asia countries form another cluster below the previous group. Remarkably, the minimum point of the GDP distribution (**Fig. 12B**, red) as well as the lowest exhaustion rates (**Fig. 12D**, blue) are characteristic features this area.
4. Conversely, countries in the West demonstrate the maximum GDP, with the peak located in the upper right corner. They solely occupy this quadrant of the elastic map, forming isolated cluster (**Fig. 12B**, red). The centroid of this cluster tends toward the highest values of the morbidity rate, *a* (**Fig. 12C**, green), the lowest values of exhaustion, K$_2$ (**Fig. 12D**, blue), as well as mobilization rate, log($q$) (**Fig. 12E**, magenta).
5. Let's highlight two additional mini-clusters, exclusively composed of countries from Africa, Asia, and Eastern Europe. These mini-clusters can be found in the lower left corner of the elastic map, exhibiting high morbidity (**Fig. 12C**, green), exhaustion rates (**Fig. 12D**, blue), but relatively low GDP values (**Fig. 12B**, red) and in the lower right corner, characterized by high morbidity (**Fig. 12C**, green), mobilization (**Fig. 12E**, magenta) rates, but low exhaustion rates (**Fig. 12D**, blue).

To visualize these results on top of a geographical map, we introduced two methods of assigning a color to the country: firstly, representing a position of the country on the 2D elastic map utilizing Ziegler 2D color encoding [47] and, secondly, 3D PCA analysis with L*a*b* encoding for color representation (**Fig. 13**). Using these color schemes is advised because in 2D and especially 3D the human visual perception of color is essentially non-euclidean so simple color scheme such as RGB can produce colors visually similar for significantly distinct coordinate values. The first method uses color to represent a position of a data point on the two-dimensional principal manifold (**Fig. 13A**). In the second case, given the original factor space's 4 dimensions, a PCA analysis was performed to reduce the dimensionality to 3 dimensions. Afterwards, positions of the data points on the first two principal components were mapped to the a*, and b* values while the third principal component was mapped onto L* component of the L*a*b* color scheme (**Fig. 13B**). Therefore, each country is assigned an individual color, representing its coordinates in these two data representations. Therefore, social communities demonstrating similar response scenarios to stress factors share similar shades. This can be compared with the colors showing the regions on the world map formed by neighbouring countries with shared traditions and cultural values. There can be exceptions where the behavior of populations from different parts of the world exhibits striking similarities. Comparison of geographical location of countries and their socio-economic characteristics and exploring the duality of countries existence in both geographical and data feature space is a standard approach in global statistical analyses in political sciences [48].

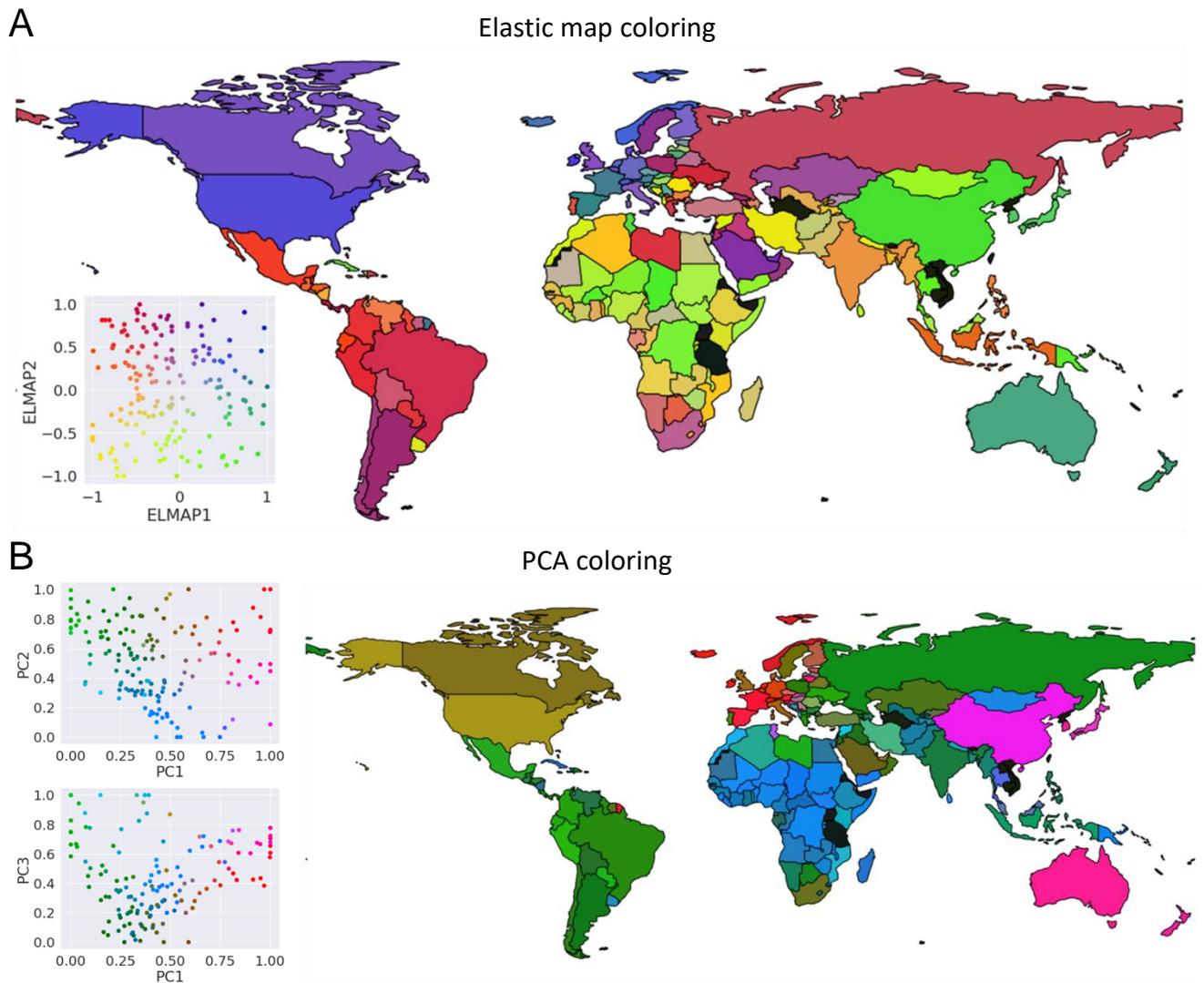

**Fig. 13. A** – Elastic map coloring utilizing Ziegler 2D color map; **B** – PCA coloring using three robust features + GDP. We apply L*a*b* color mode for the first three principal components: $PC_1$ = a, $PC_2$ = b, $PC_3$ = L. Countries not included in the analysis are marked in black.

### 4.3. Cluster analysis

In order to uncover a limited number of patterns in the distribution of model parameter values and provide explanations for the aforementioned results, we conducted a study aimed at identifying subsets of countries that form clusters based on their estimated model parameter value sets. Here, unsupervised analysis is employed to identify a limited number of characteristic scenarios or classes that depict how countries interact with the pandemic outbreak. Restricting the number of potential scenarios should facilitate the development of a global protocol for addressing future viral pandemics while recognizing the varying social reactivity specific to different societies. Clustering analysis is a standard tool for achieving this objective [49].

The use of spectral clustering and DBSCAN [50,51] methods revealed their limited effectiveness to deal with the distribution of model parameters, leading to the clustering which is difficult to interpret. Conversely, the analysis using the stabilized k-means method (with n = 1000 initial centroids) [51,52] demonstrated the potential for dividing the global dataset into clusters with interpretable composition. The optimal number of clusters derived from this method only slightly depended on the chosen evaluation measure. When calculating the adjusted Rand index (ARI), a local peak was observed with N = 6 and N = 7 clusters (**Fig. 14A**). Meanwhile, utilizing the silhouette score led to N = 7 clusters (**Fig. 14B**). Class distributions were represented as marked points on the elastic map (**Fig. 14D-F**) and on the world map, where each color corresponded to a specific class number (**Fig. 14G**).

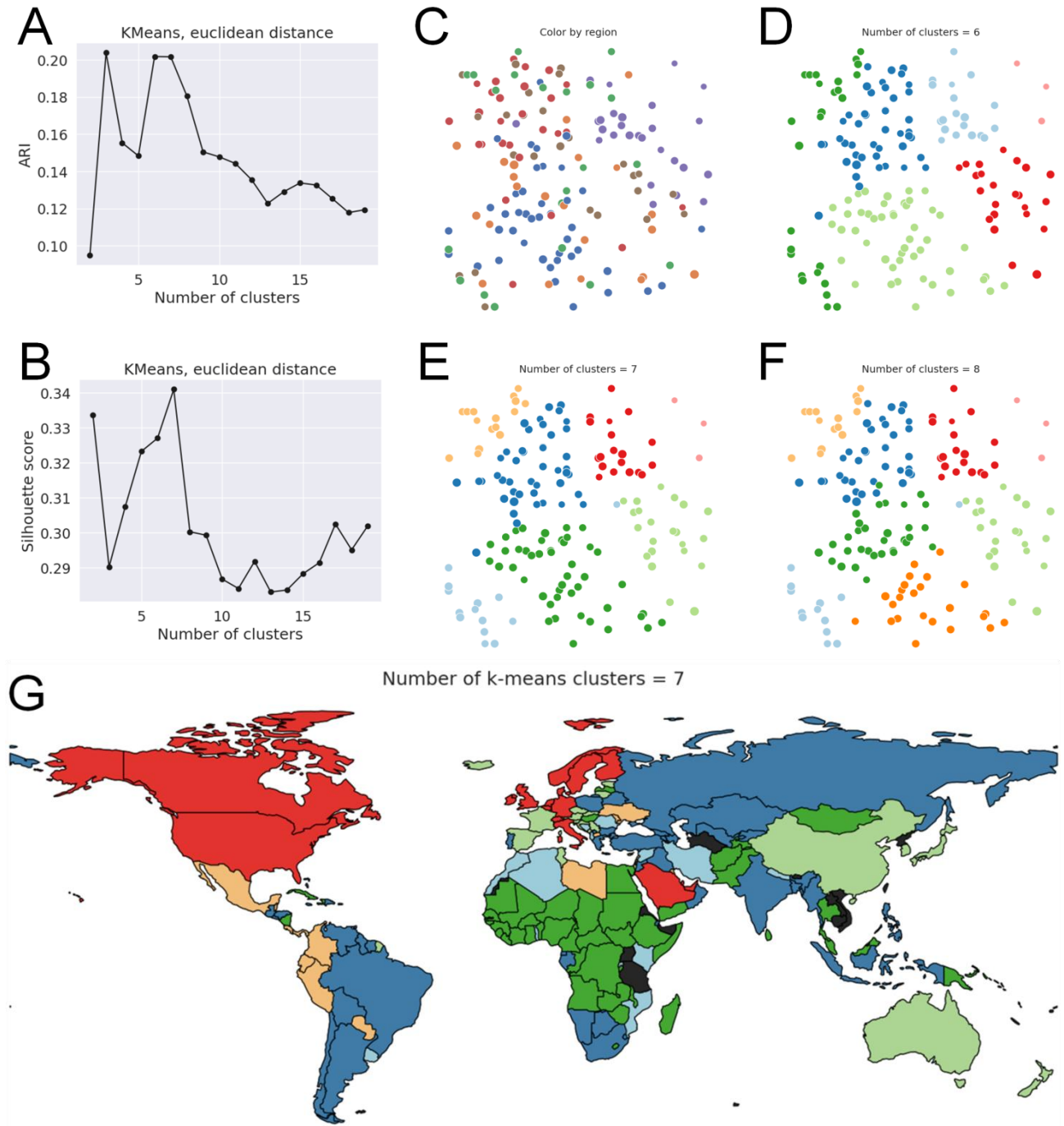

**Fig. 14.** Unsupervised data classification via stabilized k-means method (n = 1000). Countries not included in the analysis are marked in black.

The application of Ward's agglomerative hierarchical clustering yielded 5-9 and 9 clusters when using the ARI (**Fig. 15A**) and silhouette score (**Fig. 15B**) methods, respectively. It is worth noting that both clustering methods consistently grouped a few mini-groups of territories (consisting of less than 5 countries) into independent classes, even when N ≥ 6 (evident in **Fig. 14D-F** and **Fig. 15D-F**). This suggests that the true number of global clusters may not be significantly larger and likely falls within the range of approximately 5 to 7.

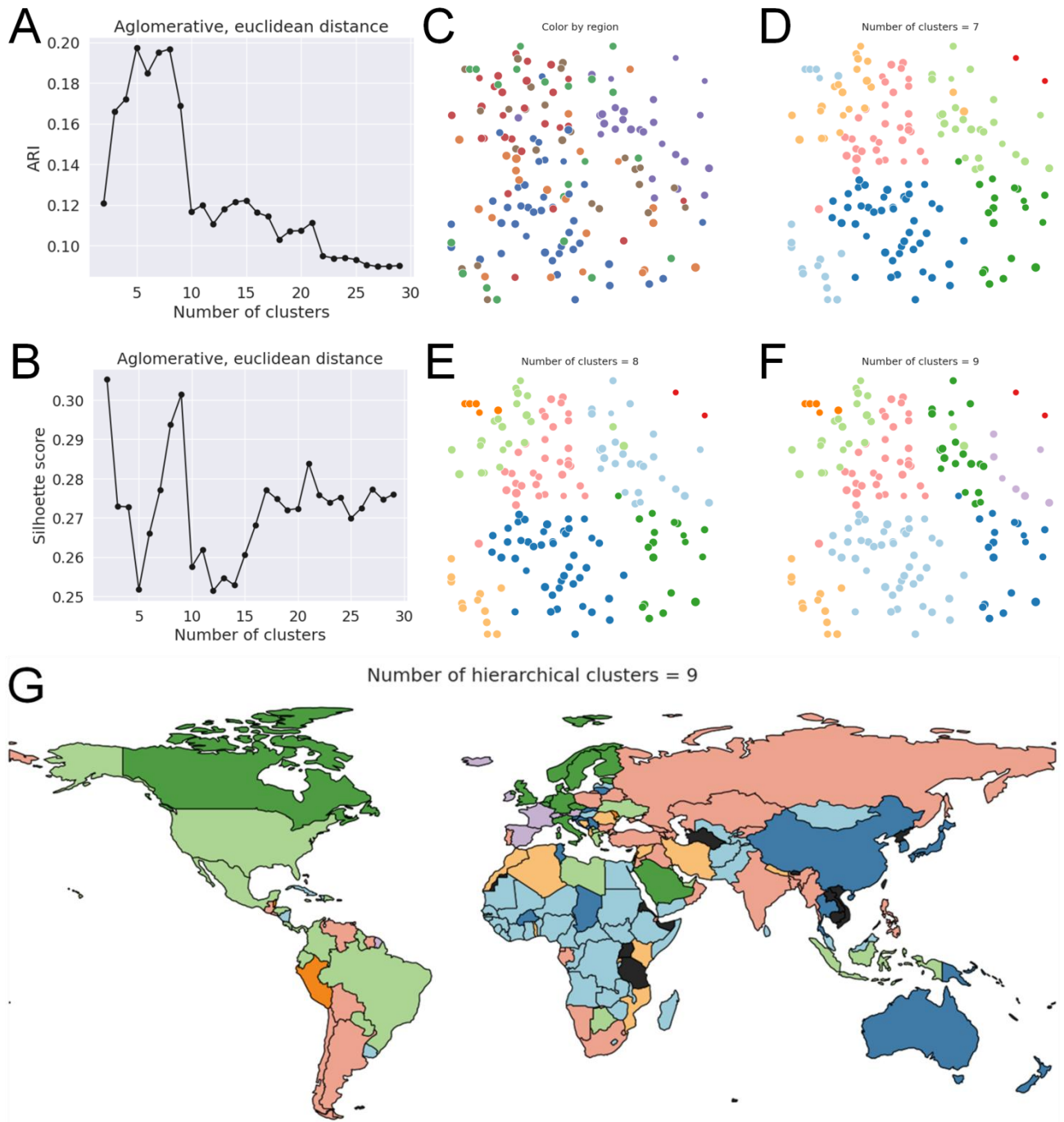

**Fig. 15.** Unsupervised data classification via agglomerative hierarchical Ward clustering. Countries not included in the analysis are marked in black.

### 4.4. Nearest neighbour graph reveals existence of archetypical scenario of adaptive COVID-19 dynamics

To further illustrate the proximity of different countries and territories in terms of their adaptive COVID-19 dynamics, we generated a graph of nearest neighbours using the kNN method (k = 2), based solely on epidemic dynamics, specifically, the coefficients $a$, $\log(q)$, and $K_2$ from **Table 1**. The comprehensive graph for all countries worldwide is provided in Appendix B: **Fig. B.1**. On these graphs, each country is connected by outgoing arrows to two and only two other countries, representing its two most similar counterparts. At the same time, the number of incoming arrows can differ from two, since a country can appear as one of the two nearest neighbours for less or more than two other countries. Reciprocal edges in such graph represent strong relationships of mutual similarity between two countries (each of them appears in the first two nearest neighbours of each other) that can reinforce their comparison.

It's worth noting that neighbouring countries are highly likely to share common economic structures, robust cross-border traffic, as well as similar historical, linguistic, cultural, traditional, and religious features. Thus, it was anticipated that in this graph, geographically proximate nations would predominantly appear adjacent or grouped into distinct "archetypes". Such simplified representation facilitates the analysis of the aforementioned features by the experts in the corresponding domain. At the same time, it should be emphasized that in such a graph, all connections beyond the closest similarity are disregarded, potentially giving the impression that some countries are completely isolated from their related nations.

As an example, consider the graph for European countries (European Union + EFTA countries + Turkey, Serbia, Israel, and United Kingdom), depicted in **Fig. 16** below. Indeed, distinct clusters united by geography are discernible. Understanding the epidemic dynamics within each country within the chosen archetype, one can observe striking similarities, as indicated by colored maps (**Fig. 13-15**).

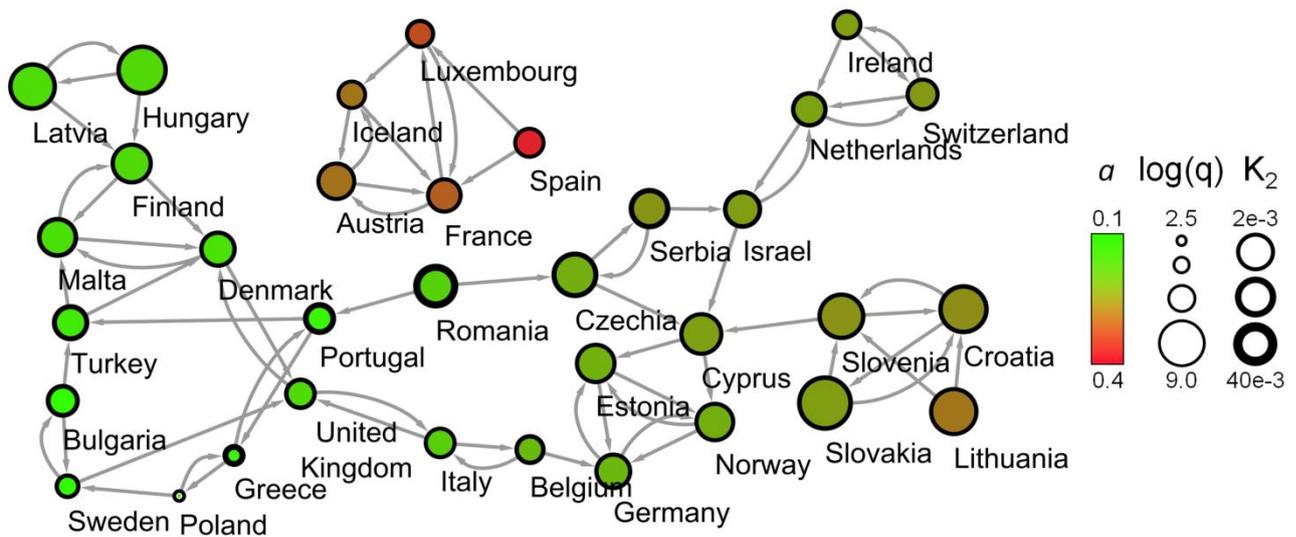

**Fig. 16.** The visualization of the nearest neighbours graph for European countries (European Union + EFTA countries + Turkey, Serbia, Israel, UK). The kNN method (k = 2), based solely on the coefficients $a$, $\log(q)$, $K_2$, characterizing epidemic dynamics, was utilized to generate the graph. Euclidean distance for the centered and scaled to unit variance parameter values was used as dissimilarity metrics. Reciprocal links represent pairs of countries that are mutually nearest neighbours.

Notable archetypes in such augmented list of European countries include the following:
1. "Particular" cluster: France-Spain-Austria-Iceland-Luxembourg.
2. Conditionally "Southern" cluster: Turkey-Bulgaria-Greece-Portugal-Malta with northern group Sweden-Denmark-Poland joining them.
3. Cluster: Latvia-Hungary-Finland.
4. "Central Eastern" cluster: Slovakia-Slovenia-Croatia-Lithuania.
5. "Northern" cluster: Estonia-Germany-Norway.
6. Cluster: Ireland-Netherlands-Switzerland.

Differing epidemic dynamics are evident within these archetypes. Notably, the "Particular" cluster features a well-expressed plateau between epidemic waves, characterized by the highest morbidity rates $a$. Meanwhile, the 3rd archetype similarly cluster represents pandemia dynamics with a lengthy plateau, with a less substantial decline in amplitude relative to the first wave's peak. Here, the parameter $a$ is low, while society's mobilization rate, $\log(q)$, is high. The "Southern" archetype, however, demonstrates lower morbidity rates, $a$, and relatively low mobilization rates, $\log(q)$, contrasting with high exhaustion rates, $K_2$.

The study of the nearest neighbour graph revealed largely anticipated archetypes. However, there are exceptions. In daily life, people's behavior is directly governed by laws and regulations, even in the context of acute interventions. As a result, the cultural characteristics of society may play a less significant role but continue to be a factor of similarity. This can explain mutually similar pairs of countries such as Finland-Malta, Netherlands-Israel, and so on.

## 5. CONCLUSIONS

In conclusion, our theoretical examination of the first wave of the epidemic using the $SIR_{SS}$ model highlights the influence of socio-cultural features and unique characteristics specific to each country. By analyzing COVID-19 statistics for all countries worldwide, we found that a few parameters can capture these characteristics and reflect societal response dynamics. Local authorities can utilize these insights to optimize strategies in combating epidemics until vaccines are available.

However, the model has limitations in incorporating multifaceted factors such as evolving medical, biological methods of protection, governmental decisions, economic trends, and viral mutations. Consequently, it requires constant modification and inclusion of numerous processes in response to these shifts, posing challenges in managing its evolving complexity. Furthermore, the $SIR_{SS}$ model solely considers official COVID-19 statistics, potentially missing unreported or latent cases, limiting its ability to accurately reflect the true infected population. This reinforces its role as an assessment tool rather than a predictive one.

Despite these limitations, the study underscores the importance of a system of adapting models that will provide us with tools for quantifying various situations and playing out various scenarios *in silico* to develop anti-epidemic strategies specific to a particular society: country, region or social group. The balance between model complexity and parameter reliability remains a major challenge, emphasizing the need for robust modeling incorporating social processes, immunity dynamics, viral evolution, and economic factors.

The analytical theory of the first wave demonstrates consistent properties in the phase plane, including insensitivity to the initial condition $I_0$ and the decrease of the first wave's peak through feedback on $I^2$. Variations in the parameters $a$, $b$, $q$, and $I_0$ do not result in qualitative changes in the phase portrait, indicating similar outbreak scenarios during the initial stage of an epidemic. Variability in observed COVID-19 statistics is largely influenced by quantitative variations in the kinetic constants specific to each country or social group.

Observing density distributions on the elastic map identifies several prominent clusters. Central and South American countries exhibit the highest density near the upper left corner (highest exhaustion rate, low GDP, low mobilization rate), while Sub-Saharan Africa predominate in the lower left quadrant (lowest GDP, low morbidity rate). Countries in the West demonstrate high GDP values and low exhaustion rate in the upper right quadrant. The clustering analysis uncovers a restricted set of 5-7 distinct modes that delineate how society responds to the onset of epidemics. This count approximately aligns with the number of economic-geographical macro-regions, highlighting several exceptions.

In summary, our analysis of the COVID-19 pandemic underscores the intertwined influence of human responses and natural occurrences, where societal reactions differ across countries with naïve populations. The findings provide valuable insights into the first surge dynamics, enabling a better understanding of societal responses and supporting the development of tailored strategies for effective control of any novel epidemics in the future.

**CRediT AUTHORSHIP CONTRIBUTION STATEMENT**

**Innokentiy Kastalskiy:** Conceptualization, Methodology, Software, Formal analysis, Investigation, Visualization, Writing – original draft, Writing – review & editing. **Andrei Zinovyev:** Methodology, Software, Formal analysis, Investigation, Visualization, Writing – original draft, Writing – review & editing. **Evgeny Mirkes:** Methodology, Validation, Formal analysis, Writing – review & editing. **Victor Kazantsev:** Supervision, Funding acquisition, Writing – review & editing. **Alexander N. Gorban:** Conceptualization, Methodology, Validation, Formal analysis, Supervision, Project administration, Writing – original draft, Writing – review & editing.

**DECLARATION OF COMPETING INTEREST**

The authors declare that they have no known competing financial interests or personal relationships that could have appeared to influence the work reported in this paper.

**DATA AND CODE AVAILABILITY**

The series of COVID-19 confirmed cases for all countries worldwide are publicly available at data repository of *Our World in Data* project: https://github.com/owid/covid-19-data/tree/master/public/data (direct link to Excel file: https://covid.ourworldindata.org/data/owid-covid-data.xlsx, date of access: May 10, 2021). Raw data come from the COVID-19 Data Repository by the Center for Systems Science and Engineering (CSSE) at Johns Hopkins University (JHU) at https://github.com/CSSEGISandData/COVID-19.

Excel file with Table 1, full size versions of the figures provided in Appendix A, as well as code used to produce the results presented herein can be found in public GitHub repository https://github.com/lamhda/COVID_SIRss.

**ACKNOWLEDGMENTS**

The work of I. Kastalskiy and V. Kazantsev was funded by the Ministry of Science and Higher Education of the Russian Federation by the federal academic leadership program "Priority 2030".

# APPENDIX A

### Epidemic trajectories of the SIR$_{SS}$ model and their comparison with COVID statistics for all countries of the world analyzed

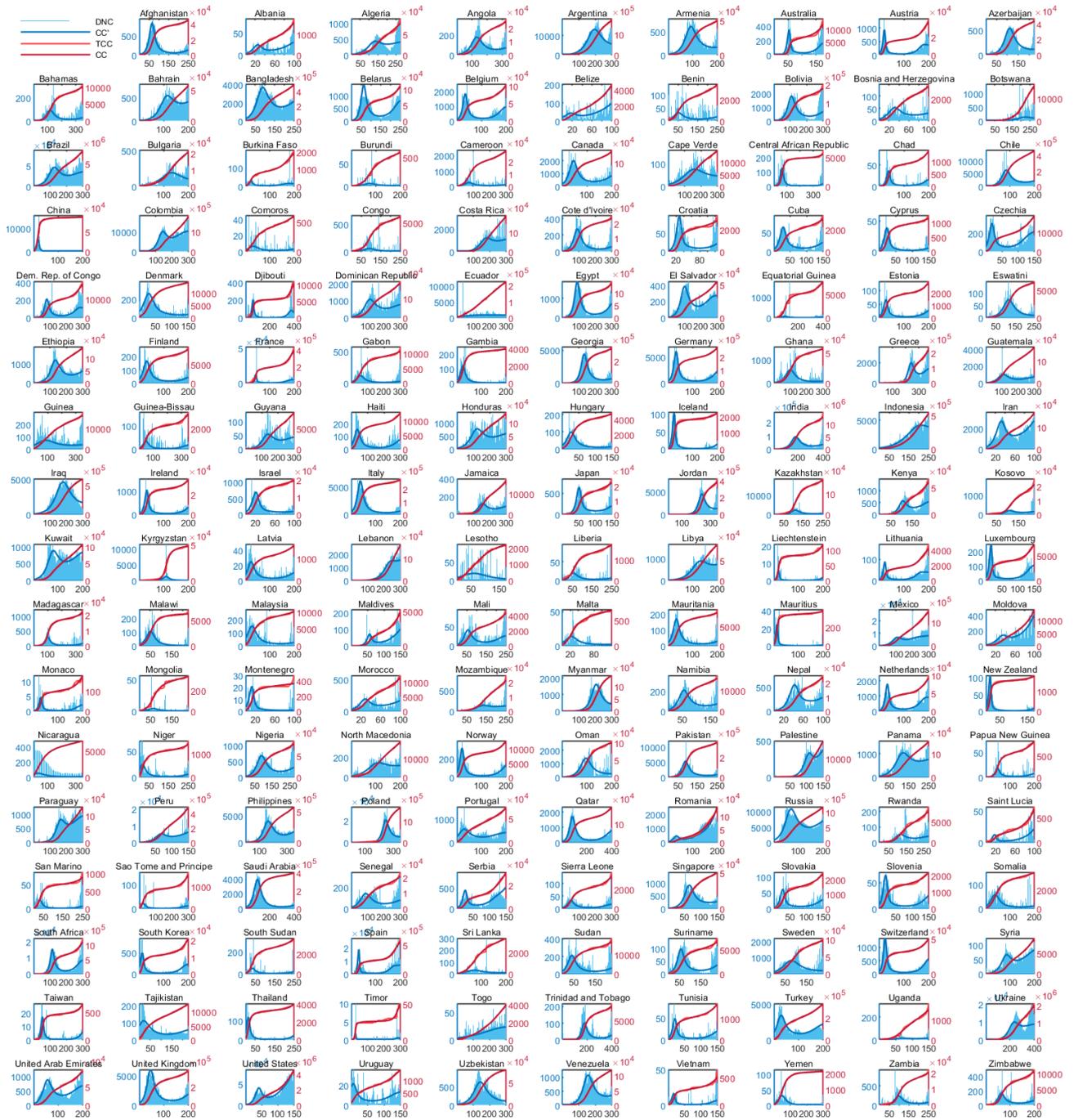

**Fig. A.1.** Fitted COVID data in absolute values: total confirmed cases (TCC, scarlet) and daily new cases (DNC, light blue stem plot) compared with cumulative cases (CC = I + R, dark red) and its derivative (CC', dark blue) multiplied on population vs. N$_{days}$.

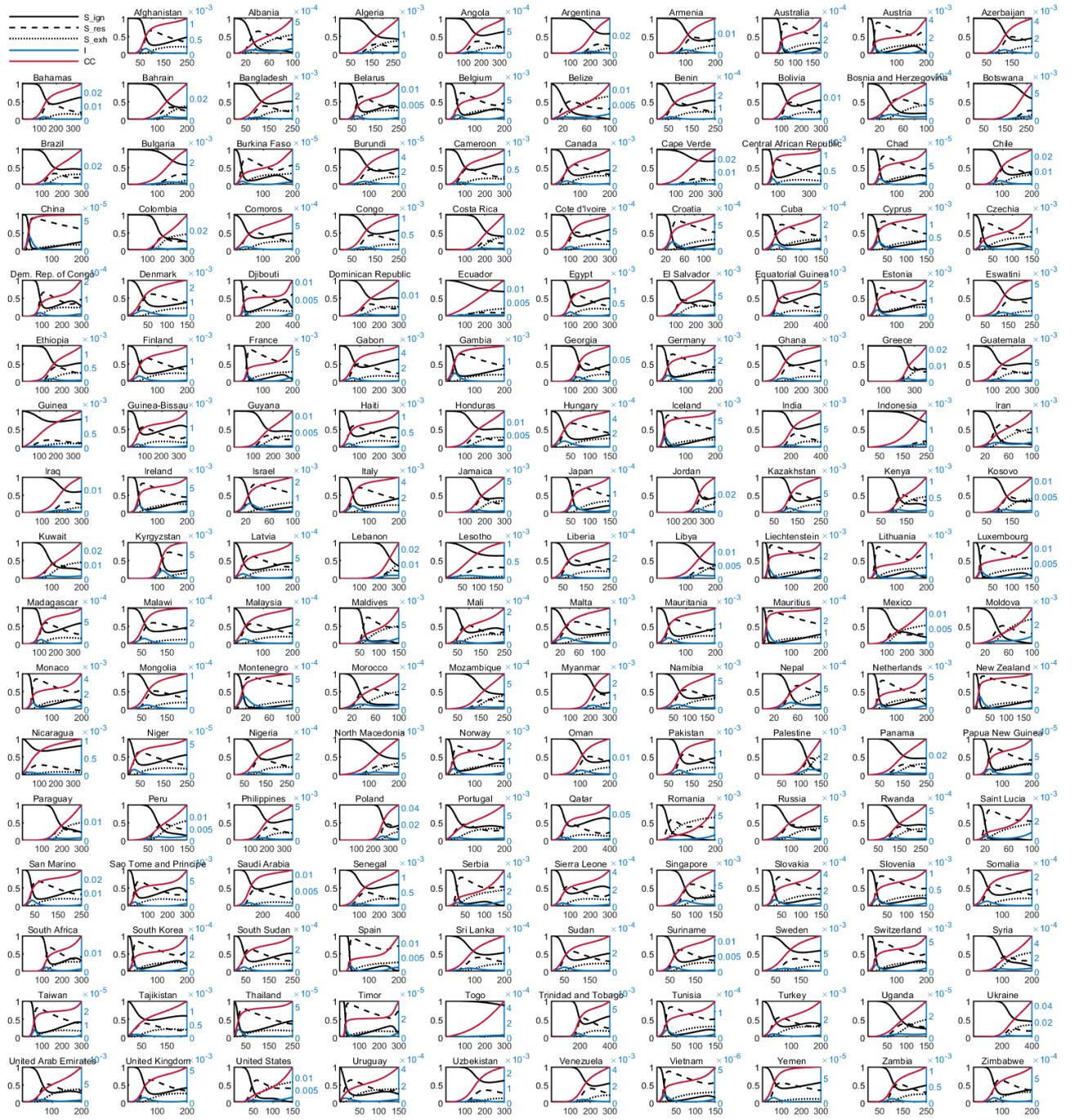

**Fig. A.2.** The dynamics of the $SIR_{SS}$ model: $S_{ign}$ (solid black), $S_{res}$ (dashed black), $S_{exh}$ (dotted black), I (blue), and cumulative cases CC = I + R (red) vs. $N_{days}$.

# APPENDIX B

## The comprehensive nearest neighbours graph for all countries of the world analyzed

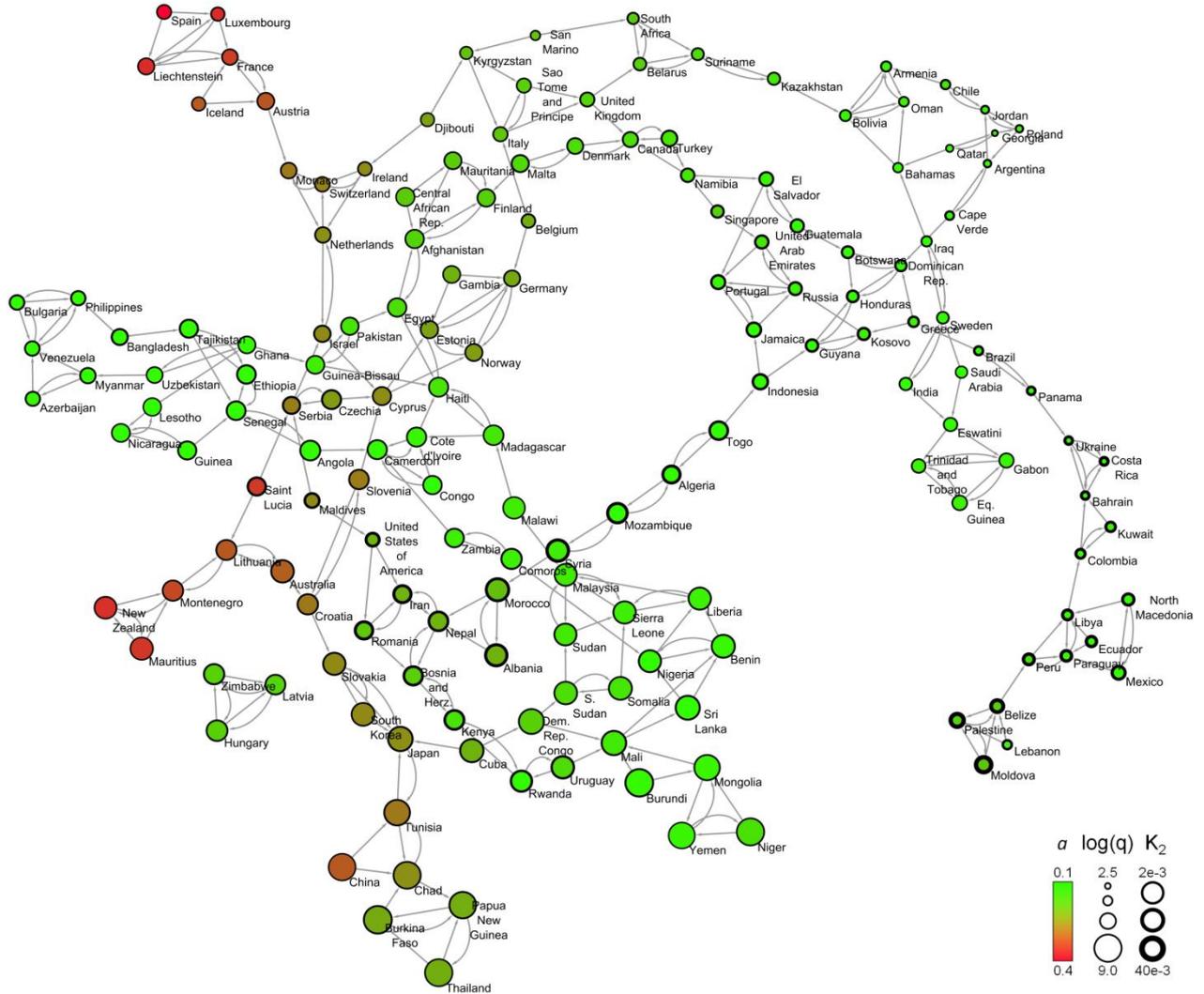

**Fig. B.1.** The nearest neighbours graph for all countries worldwide. The kNN method (k = 2) was utilized to generate it. The graph is based solely on the coefficients $a$, $\log(q)$, and $K_2$, characterizing epidemic dynamics. Euclidean distance for the centered and scaled to unit variance parameter values was used as dissimilarity metrics. Reciprocal links represent pairs of countries that are mutually nearest neighbours.

Highlights

# Exploring the impact of social stress on the adaptive dynamics of COVID-19: Typing the behavior of naïve populations faced with epidemics

Innokentiy Kastalskiy, Andrei Zinovyev, Evgeny Mirkes, Victor Kazantsev, and Alexander N. Gorban

- The kinetics of epidemics is strongly coupled with the dynamics of social stress
- We developed a system of models that combines the dynamics of epidemics and social stress
- Already a simple model of this family accurately describes the dynamics of the first wave of COVID-19
- The kinetics of the first wave of COVID-19 and social stress were analyzed for 169 countries
- The main clusters of behavioral response of different countries are identified and presented on a map